\documentstyle[preprint,aps]{revtex}
\begin{document}
 \tightenlines
 \preprint{Monograph CBPF-MO-03-01 (December/2001)}
\title{THE QUANTUM MECHANICS SUSY ALGEBRA:
AN INTRODUCTORY REVIEW}
\author{R. de Lima Rodrigues\thanks{Departamento de
Ci\^encias Exatas e da Natureza, Universidade Federal de Campina Grande,
Cajazeiras - PB, 58.900-000, Brazil. E-mail: rafaelr@cbpf.br or
rafael@fisica.ufpb.br}
\\
Centro Brasileiro de Pesquisas F\'\i sicas\\
Rua Dr. Xavier Sigaud, 150\\
CEP 22290-180, Rio de Janeiro-RJ, Brazil}

\maketitle

\begin{abstract}
Starting  with the Lagrangian formalism with $N=2$ supersymmetry in terms of
two Grassmann variables  in Classical Mechanics, the Dirac canonical
quantization method is implemented.
The $N=2$ supersymmetry algebra  is associated to one-component and
two-component eigenfunctions considered in the Schr\"odinger picture of
Nonrelativistic Quantum Mechanics.  Applications are contemplated.
\end{abstract}

\vspace{0.5cm} PACS numbers: 11.30.Pb, 03.65.Fd, 11.10.Ef

\pacs
\newpage

\section{INTRODUCTION}

We present a review work considering the Lagrangian formalism for the
construction of one dimension supersymmetric (SUSY) quantum mechanics (QM)
with $N=2$ supersymmetry (SUSY) in a non-relativistic context. In this
paper, the supersymmetry with two Grassmann variables ($N=2$) in classical
mechanics is used to implement the Dirac canonical quantization method and
the main characteristics of the SUSY QM  is considered in detail. A general
review on the SUSY algebra in quantum mechanics and the procedure on like
to build a SUSY Hamiltonian hierarchy in order of a complete spectral
resolution it is explicitly applied for the  P\"oschl-Teller potential I.
We will follow a more detailed discussion for the case of this problem
presents unbroken SUSY and broken SUSY. We have include a large number of
references where the SUSY QM works, with emphasis on the one-component
eigenfunction under non-relativistic context. But we indicate some articles
on the SUSY QM from Dirac equation of relativistic quantum mechanics. The
aim of this paper is to stress the discussion how arise and to bring out
the correspondence between SUSY and factorization method in quantum
mechanics. A brief account of a new scenario on SUSY QM to two-component
eigenfunctions, makes up the last part of this review work.

SUSY first appeared in field theories in terms of bosonic and fermionic
fields\footnote{A bosonic field (associated with particles of integral or
null spin) is one particular case obeying the Bose-Einstein statistic and a
fermionic field (associated to particles with semi-integral spin) is that
obey the Fermi-Dirac statistic.}, and the possibility was early observed
that it can accommodate a Grand-Unified Theory (GUT) for the four basic
interactions of Nature (strong, weak, electromagnetic and gravitational)
\cite{Swieca93}. The first work on the superalgebra in the space-time
within the framework of the Poincar\'e algebra was investigated by Gol'fand
and Likhtman \cite{GL71}. On the other hand, Volkov-Akulov have considered
a non-renormalizable realization of supersymmetry in field theory
\cite{VA72}, and Wess-Zumino have presented a renormalizable supersymmetric
field theory model \cite{WZ74}.

 Recently the SUSY QM has
also been investigated with pedagogical purpose in some booktexts
\cite{textqm} on quantum mechanics giving its connections with the
factorization method \cite{Inf}. Starting from factorization method new
class of one-parameter family of isospectral potential
in one dimension has been constructed
with the energy spectrum coincident with that of the harmonic oscillator by
Mielnik \cite{Miel}.
In recent literature, there are some interesting books on
supersymmetric classical and quantum mechanics emphasizing different
approach and applications of the theory \cite{bjd}.

Fernandez {\it et al.} have considered the connection between factorization
method and generation of solvable potentials \cite{fernandez84}. The SUSY
algebra in quantum mechanics initiated with the work of Nicolai \cite{N}
and elegantly formulated by Witten \cite{W}, has attracted interest and
found many applications in order to construct the spectral resolution of
solvable potentials in various fields of physics. However, in this work,
SUSY $N=2$ in classical mechanics
\cite{Galv80,Salo82,azca82,azca83,azca86,stoya87,junker95} in a
non-relativistic scenario is considered using the Grassmann variables
\cite{Bere}. Recently, we have shown that the $N=1$ SUSY in classical
mechanics depending on a single commuting supercoordinate exists only for
the free case \cite{RWI}.

Nieto has shown that the generalized factorization observed by Mielnik
\cite{Miel} allow us to do the connection between SUSY QM  and the inverse
method \cite{Nieto84,suku85inv}.  The first technique that have been used
to construct some families of isospectral order second differential
operators is based on a theorem due to Darboux, in 1882 \cite{darboux}.

 J. W van Holten {\it et
al.} have written a number of  papers dealing with SUSY mechanical systems
\cite{van82,van83,van84,macfar84,van93a,van93,van95a,van95,van96,van01}.
The canonical quantization of $N=2$ (at the time called $N=1$) SUSY models
on spheres and hyperboloids \cite{van83} and on arbitrary Riemannian
manifolds have been considered in \cite{van84}; its $N=4$ (at that time
called $N=2$) generalization is found in \cite{macfar84}; SUSY QM in
Schwarzchild background was studied in \cite{van93a}; New so-called
Killing-Yano supersymmetries were found and studied in
\cite{van93,van95a,van95}; General multiplet calculus  for locally
supersymmetric point particle models was constructed in \cite{van96} and
the relativistic and supersymmetric theory of fluid mechanics in 3+1
diemnsions has been investigated by Nyawelo-van Holten \cite{van01}. The
vorticity in the hydrodynamics theory  is generated by the fermion fields
\cite{JC00}.

D'Hoker and Vinet have also written a number
of  papers dealing with classical and quantum
mechanical supersymmetric Lagrangian mechanical systems.
They have shown
that a non-relativistic spin $\frac 12$ particle
in the field of a Dirac magnetic monopole exhibits a large SUSY
invariance \cite{Eric84}. Later,  they have published some other
interesting works on the construction of conformal superpotentials
for a spin $\frac 12$ particle in the field of a Dyon and the magnetic
monopole
and $\frac{1}{r^2}-$potential for particles in a Coulomb potential
\cite{Eric85}.
However, the supersymmetrizatin of the action for the charge monopole
system have been also developed by Balachandran {\it et al.} \cite{Bala80}.

A new SUSY QM system given by a non-relativistic charged spin-$\frac 12$
particle in an extended external electromagnetic field was obtained by
Dias-Helayel \cite{dias-helayel01}.

Using a general formalism for the non-linear quantum-mechanical $\sigma$
model, a mechanism of spontaneous breaking of the supersymmetry at the
quantum level related to the uncertainty of the operator ordering has been
obtained by Akulov-Pashnev \cite{akulov85}. In \cite{akulov85} is noted the
simplicity of the supersymmetric $O(3)$-or $O(2,1)$-ivariant Lagrangian
deduced there when  compared with the analogous obtained using real
superfields \cite{van84}. The mechanism of spontaneous breaking of the
supersymmetry in quantum mechanics has also been investigated by Fuchs
\cite{fuchs86}.

Barcelos and others have implemented the Dirac quantization method in
superspace and found the SUSY Hamiltonian operator \cite{Barce86}.
Recently, Barcelos-Neto and Oliveira have investigated the transformations
of second-class into first-class constraints in supersymmetric classical
theories for the superpoint \cite{Barce97}. Junker-Matthiesen have also
considered the Dirac's canonical quantization method for the
non-relativistic superpaticle \cite{junker94}. In the interest of setting
an accurate historical record of the subject, we point out that, by using
the Dirac's procedure for two-dimensional supersymmetric non-linear
$\sigma$-model, Eq. (13) of the paper by Corrigan-Zachos \cite{Zac79} works
certainly for a SUSY system in classical mechanics.

A generalized Berezin integral and fractional superspace measure arise as a
deformed $q$-calculus is developed on the basis of an algebraic structure
involving graded brackets.  In such a construction of fractional
supersymmetry the $q$-deformed bosons play a role exactly analogous to that
of the fermions in the familiar supersymmetric case, so that the SUSY is
identified as translational invariance along the braided line by Dunne {\it
et al.} \cite{dune-a96}. An explicit formula has been given in the case of
real generalised Grassmann variable, $\Theta^n = 0$, for arbitrary integer
$n = 2, 3, \cdots$ for the transformations that leave the theory invariant,
and it is shown that these transformations possess interesting group
properties by Azc\'arraga and Macfarlane  \cite{mac-a96}. Based on the idea
of quantum groups \cite{monteiro94} and paragrassmann variables in the
$q$-superspace, where $\Theta^3 = 0,$ a generalization of supersymmetric
classical mechanics with a deformation parameter $q= \exp{\frac{2 \pi
i}{k}}$ dealing with the $k =3$ case has been considered by  Matheus-Valle
and Colatto \cite{colatto96}.

 The reader can find a large number of studies
of fractional supersymmetry in literature. For example, a new geometric
interpretation of SUSY, which apllies equally in the fractional case.
Indeed, by means of a chain rule expansion, the left and right derivatives
are identified with the charge $Q$ and covariant derivative $D$ encountered
in ordinary/fractional supersymmetry and this leads to new results for
these operators \cite{dune-a97}.

 Supersymmetric
Quantum Mechanics is of intrinsic mathematical interest in its own as it
connects otherwise apparently unrelated (Cooper and Freedman \cite{CF})
second-order differential equations.

For a class of the dynamically broken supersymmetric quantum-mechanical
models proposed by Witten \cite{W}, various methods of estimating the
ground-state energy, including  the instanton developed by Salomonson-van
Holten \cite{Salo82} have been examined by Abbott-Zakrzewski \cite{A}. The
factorization method \cite{Inf} was generalized by Gendenstein \cite{Gend}
in context of SUSY QM in terms of the reparametrization of potential in
which ensure us if the resolution spectral is achieved by an algebraic
method. Such a reparametrization between the supersymmetric potential pair
is called shape invariance condition.

In the Witten's model  of SUSY QM the Hamiltonian of a certain quantum
system is represented by a pair $H_\pm,$ for which all energy levels except
possibly the ground energy eigenvalue are doubly degenerate for both
$H_\pm.$  As an application of the simplest
of the graded Lie algebras of the supersymmetric fields theories, the SUSY
Quantum Mechanics embodies the essential features of a theory of
supersymmetry, i.e., a symmetry that generates transformations between
bosons and fermions or rather between bosonic and the fermionic sectors
associated with a SUSY Hamiltonian. SUSY QM is defined (Crombrugghe and
Rittenberg \cite{Crom} and  Lancaster \cite{Lan})
 by a graded Lie algebra satisfied by the charge
 operators $Q_{i}
(i= 1,2,\ldots, N)$ and the SUSY Hamiltonian $H$.
The $\sigma$ model and supersymmetric gauge theories
have been investigated in the context of SUSY QM by Shifman {\it et al.}
\cite{Shif88}.

While in field theory one works with SUSY as being a symmetry associated
with transformations between bosonic and fermionic particles. In this case
one has transformations between the component fields whose intrinsic spin
differ by $\frac 12\hbar.$ The energy of potential models of the SUSY in
field theory  is always positive semi-defined
\cite{Swieca93,Salam,Gates83,freund86,p-wess86,prem86,azca87,azca88,burges83}.
Here is the main difference of SUSY between field theory and quantum
mechanics. Indeed, due to the energy scale to be of arbitrary origin the
energy in quantum mechanics is not always positive.

Using supergraph metods, Helayel-Neto {\it et al.} have derived
the chiral and antichiral superpropagators \cite{helayel86};
have calculated
the chiral and gauge anomalies for the supersymmetric Schwinger model
\cite{helayel87}; under certain asumption on the torsion-like
explicitly breaking term,
one-loop finiteness without spoilong the Ricci-flatness of
the target manifold \cite{helayel87b}.

 After a considerable  number of works investigating SUSY in Field Theory,
confirmation of SUSY as high-energy unification theory is missing.
Furthermore, there exist phenomenological applications of the $N=2$ SUSY
technique in quantum mechanics \cite{feno}.

 The SUSY hierarchical prescription \cite{ABI} was
utilized by Sukumar \cite{Suku85} to solve  the energy spectrum of the
P\"oschl-Teller potential I (PTPI). We will use their notation.

The two first review work on SUSY QM with various applications were reported
 by Gendenshtein-Krive and Haymaker-Rau \cite{GR} but does they not consider
the Sukumar's method \cite{Suku85}.
In next year to the review work by Gendenshtein-Krive,
Gozi implemented an approach on the nodal structure of
supersymmetric wave functions
\cite{G} and Imbo-Sukhatme have investigated the conditions
for nondegeneracy in supersymmetric quantum mechanics \cite{Sukhat86}.

In the third in a series of papers dealing with
families of isospectral Hamiltonians, Pursey
has been used the theory of isometric operators
to construct a unified treatment of three procedures
existing in literature for generating one-parameter families of
isospectral Hamiltonian \cite{Pur86}.
In the same year, Casta\~nos {\it et al.} have also shown that
any n-dimensional scalar
Hamiltonian possesses hidden supersymmetry provided its spectrum is
bounded from below \cite{Casta86}.

Lahiri {\it et al.} have investigated the transformation considered by
Haymaker-Rau \cite{GR}, viz., of the type $x=\ell n y$ so that the radial
Schr\"odinger equation for the Coulomb potential becomes a unidimensional
Morse-Schr\"odinger equation, and have stablished a procedure for
constructing the SUSY transformations \cite{lahiri87}.

Cooper-Ginocchio have used the Sukumar's method \cite{Suku85} gave strong
evidence that the more general Natanzon potential class not shape invariant
and found the PTPI as particular case \cite{CGK}. In the works of
Gendenshtein \cite{Gend} and Dutt {\it et al.} \cite{Sukhat88} only the
energy spectrum of  the PTPI was also obtained but not the excited state
wave functions. The unsymmetric case has been treated algebraically by
Barut, Inomata and Wilson \cite{BIW}. However in these analysis only
quantized values of the coupling constants of the PTPI have been obtained.

 Roy-Roychoudhury have shown that the
finite-temperature effect causes spontaneous breaking of SUSY QM, based on
a superpotential with (non-singular) non-polynomial character, and
Casahorran has investigated the superymmetric Bogomol'nyi bounds at finite
temperature \cite{Roy88}.

Jost functions are studied within framework of SUSY QM by
Talukdar {\it et al.}, so that it is seen that some of the existing results
follow from their work in a rather way \cite{Talukdar89}.

Instanton-type quantum fluctuations in supersymmetric quantum mechanical
systems with a double-well potential and a tripe-well potential have been
discussed by Kaul-Mizrachi \cite{KM} and the ground state energy was found
via a different method considered by Salomonson-van Holten \cite{Salo82}
and Abbott-Zakrzewski \cite{A}.

Stahlhofen showed that the shape invariance condition \cite{Gend}
for supersymmetric potentials and the factorization condition for
Sturm-Liouville eigenvalue problems are equivalent \cite{Sta}.
Fred Cooper {\it et al.} starting from  shape invariant potentials
\cite{Gend} applied an operator transformation for the
P\"oschl-Teller potential I and found that the Natanzon class of
solvable potentials \cite{FGW}. The supersymmetric potential
partner pair through the Fokker-Planck superpotential has been
used to deduce the computation of the activation rate in
one-dimensional bistable potentials to a variational calculation
for the ground state level of a non-stable quantum system
\cite{Mar89}.

A systematic procedure using SUSY QM has been presented for calculating the
accurate energy eigenvalues of the Schr\"odinger equation that obviates the
introduction of large-order determinants by Fernandez, Desmet and Tipping
\cite{FDT}.

SUSY has also been applied for Quantum Optics. For instant, let us point out
that the superalgebra of the Jaynes-Cummings model is described and the
presence of a gap in the energy spectrum indicates a
spontaneous SUSY breaking. If the gap tends to zero the SUSY is restored
\cite{Andreev89}. In another work, the Jaynes-Cummings model
for a two-level atom interacting with an electromagnetic field
is analyzed in terms of SUSY QM and
their eigenfunctions are deduced \cite{Lee90}.
Other applications on the SUSY QM
to Quamtum Optics can be found in \cite{optics}.

Mathur has shown that the symmetries of the Wess-Zumino model put severe
constraints on the eigenstates of the SUSY Hamiltonian simplifying the
solutions of the equation associated with the annihilation conditions for
a particular superpotential \cite{Mat90}. In this interesting work, he has
found the non-zero energy spectrum and all excited states are at least
eightfold degenerate.

The connection of the PTPI with new isospectral
potentials has been studied by Drigo Filho \cite{Elso90}.
Some remarks on a new scenario of SUSY QM by imposing a
structure on the raising and lowering operators have been found for
the 1-component eigenfunctions \cite{Janu}.
The unidimensional SUSY oscillator has been used to construct the
strong-coupling limit of the Jaynes-Cummings model exhibiting a noncompact
ortosymplectic SUSY by Shimitt-Mufti \cite{SM90}.

The propagators for shape invariant potentials and certain recursion
relations for them both in the operator formulation as well as in the path
integrals were investigated with some examples by Das-Huang \cite{DH}.

At third paper about a review on SUSY QM, the key ingredients on the
quantization of the systems with anticommuting variables and supersymmetric
Hamiltonian was constructed by emphasizing the role of partner potentials and
the superpotentials have been discussed by Lahiri, Roy and Bagchi
\cite{Lahi}. In which  Sukumar's supersymmetric procedure was applied for
the following potentials: unidimensional harmonic oscillator, Morse
potential and $sech^2x$ potential.

The formalism of SUSY QM has also been used to realize Wigner
superoscillators in order to solve the Schr\"odinger equation for the
isotonic oscillator (Calogero interaction) and radial oscillator
\cite{Jaya}.

Freedman-Mende considered the application in supersymmetric quantum
mechanics for an exactly soluble N-particle system with the Calogero
interaction \cite{FM}. The SUSY QM formalism  associated with 1-component
eigenfunctions was also applied to a planar physical system in the momentum
representation via its connection with a PTPI system. There, such a system
considered was a neutron in an external magnetic field \cite{Voro}.

A supersymmetric generalization of a known solvable quantum mechanical
model of particles with Calogero interactions, with  combined harmonic and
repulsive forces have investigated by Freedman-Mende \cite{FM} and the
explicit solution for such a supersymmetric Calogero were constructed by
Brink {\it et al.} \cite{Brink}.

 Dutt {\it et al.} have investigated the PTPI system
with broken SUSY and new exactly solvable Hamiltonians via
shape invariance procedure \cite{Sukhat93}.

The formulation of higher-derivative supersymmetry  and its connection with
the Witten index has been proposed by Andrianov {\it et al.} \cite{AIS} and
Beckers-Debergh \cite{BD} have discussed a possible extension of the
super-realization of the Wigner quantization procedure considered by
Jayaraman-Rodrigues \cite{Jaya}. In \cite{BD} has been proposed a
construction that was called of a parastatistical hydrogen atom which is a
supersymmetric system but is not a Wigner system.
Results of such investigations and also of the pursuit of the
current encouraging indications to extend the present
formalism for Calogero interactions
will be reported separately.

In another work on SUSY in the non-relativistic hydrogen atom,
Tangerman-Tjon have stressed the fact that no extra particles are needed to
generate the supercharges of $N=2$ SUSY algebra when we use the spin
degrees of freedom of the electron \cite{Tan93}. Boya {\it et al.} have
considered the SUSY QM approach from geometric motion on arbitrary rank-one
Riemannian symmetric spaces via Jost functions and the Laplace-Beltrami
operator \cite{boya93}.

 In next year, Jayaraman-Rodrigues
have also identified the free parameter of the Celka-Hussin's model with
the Wigner parameter \cite{Jaya} of a related super-realized general 3D
Wigner oscillator system satisfying a super generalized quantum commutation
relation of the $\sigma_3$-deformed Heisenberg algebra \cite{JR94}. In this
same year, P. Roy  has studied the possibility of contact interaction of
anyons within the framework of two-particle SUSY QM model \cite{roy94};
indeed, at other works the anyons have been studied within the framework of
supersymmetry \cite{sen92}.

In stance in the literature, there exist four excellent review articles
about SUSY in non-relativistic quantum mechanics \cite{GR,Lahi,Fred}.
Recently the standard SUSY
formalism was also applied   for a neutron in interaction with a
static magnetic field in the coordinate representation \cite{VGM}
and the SUSY QM in higher dimensional was discussed by Das-Pernice \cite{DP}.

Actually it is well known that the SUSY QM formalism is intrinsically bound
with the theory of Riccati equation. Dutt {\it et al.} have ilusted the
ideia of SUSY QM and shape invariance conditions can be used to obtain
exact solutions of noncentral but separable potentials in an algebraic
fashion \cite{Sukhat97}. A procedure for obtaining the complete energy
spectrum from the Riccati equation has been illustrated by detailed
analysis of several examples by Haley \cite{Hale}.

Including not only formal mathematical objects and schemes but also new
physics, many different physical topics are considered by the SUSY
technique (localization, mesoscopics, quantum chaos, quantum Hall effect,
etc.) and each section begins with an extended introduction to the
corresponding physics. Various aspects of SUSY may limit themselves to
reading the chapter on supermathematics, in a book written by Efetov
\cite{Efet97}.

SUSY QM of higher order have been by Fernandez {\it et al.}
\cite{fernandez97}. Starting from SUSY QM, Junker-Roy \cite{Junk},
presented a rather general method for the construction of so-called
conditionally exactly sovable potentials \cite{Cond95}. A new SUSY method
for the generation of quasi-exactly solvable potentials with two known
eigenstates has been proposed by Tkachuk \cite{Tkac98}.

 Recently
Rosas-Ortiz has shown a set of factorization energies generalizing the
choice made for the Infeld-Hull \cite{Inf} and Mielnik \cite{Miel}
factorizations of the hydrogen-like potentials \cite{Rosa98}. The SUSY
technique has also been used to generate families of isospectral potentials
and isospectral effective-mass variations, which may be of interest, e.g.,
in the design of semiconductor quantum wells \cite{Mila99}.

The soliton solutions have been investigated for field equations defined in
a space-time of dimension equal to or higher than 1+1. The kink of a field
theory is an example of a soliton in 1+1 dimensions
\cite{Jac,Raja,Cole,Bala,Weinberg}. In this work we  consider the
Bogomol'nyi \cite{Bogo} and Prasad-Sommerfield \cite{PS} (BPS) classical
soliton (defect) solutions. Recently, from $N=1$ supersymmetric solitons
the connection between SUSY QM and the sphaleron and kinks has been
established for relativistic systems of a real scalar field
\cite{Kumar87,Boya,Aragao89,R95,Junker97,rebhan97,P99,nieuwen01}.

The shape-invariance conditions in SUSY \cite{Gend} have been generalized
for systems described by two-component wave functions \cite{Ta}, and a
two-by-two matrix superpotential associated to the linear classical
stability from the static solutions for a system of two coupled real scalar
fields in (1+1)-dimensions have been found
\cite{RV96,RPV98,Shif99,RWPV,ger01}. In Ref. \cite{Shif99} has been shown
that the classical central charge, equal to the jump of the superpotential
in two-dimensional models with minimal SUSY, is additionally modified by a
quantum anomaly, which is an anomalous term proportional to the second
derivative of the superpotential. Indeed, one can consider an analysis of
the anomaly in supersymmetric theories with two coupled real scalar fields
\cite{R99} as reported in the work of Shifman {\it et al.} \cite{Shif99}.
Besides, the stability equation for a Q-ball in 1 dimension has also been
related to the SUSY QM \cite{Mac01}.

A systematic and critical examination,  reveals that when carefully done,
SUSY is manifest even for the singular quantum mechanical models when the
regularization parameter is removed \cite{DP99}. The  Witten's SUSY
formulation for Hamiltonian systems to also a system of annihilation
operator eigenvalue equations associated with the SUSY singular oscillator,
which, as was shown, define SUSY canonical supercoherent states containing
mixtures of both pure bosonic and pure fermionic counterparts have been
extended \cite{JR99}. Also, Fernandez {\it et al.} have investigated the
coherent states for SUSY partners of the oscillator \cite{fernandez99}, and
Kinani-Daoud have built the coherent states for the P\"oschl-Teller
potential \cite{kinani01}.

In the first work in Ref. \cite{Mik00}, Plyushchay has used arguments of
minimal bosonization of SUSY QM
and R-deformed Heisenberg algebra in order to get in the second paper
in the same Ref. a super-realization for the ladder operators of the
Wigner oscillator \cite{Jaya}. While Jayaraman and Rodrigues,
in Ref. \cite{Jaya}, adopt a super-realization of the Wigner-Heisenberg
algebra ($\sigma_3-$deformed Heisenberg algebra)
as effective spectral resolution for the two-particle Calogero
interaction or isotonic oscillator, in Ref. \cite{Mik00},
using the same super-realization, Plyushchay
showed how a simple modification of the
classical model underlying Witten SUSY QM results in appearance
of $N=1$ holomorphic non-linear supersymmetry.

In the context of the symmetry of the fermion-monopole system
\cite{Eric84}, Pluyschay has shown that this system possesses
$N=\frac 32$ nonlinear supersymmetry \cite{Mik00b}.
The spectral problem of the 2D system with the quadratic magnetic field is
equivalent to that of the 1D quasi-exactly solvable systems with the sextic
potential, and the relation of the 2D holomorphic n-supersymmetry
to the non-holomorphic N-fold supersymmetry has been investigated
\cite{Mik01}.

In \cite{SMik01}, it was shown that the problem of quantum anomaly
can be resolved for some
special class of exactly solvable and quasi-exactly solvable systems.
So, in this paper it was discovered that the nonlinear supersymmetry
is related with quasi-exact solvability.
Besides, in this paper it was observed that the quantum anomaly happens
also in the case of the linear quantum mechanics and that the usual
holomorphic-like form of SUSYQM
(in terms of the holomorphic-like operators $W(x)\pm i\frac{d}{dx}$)
is special: it is anomaly free.

Macfarlane \cite{Mac97} and Azc\'arraga-Macfarlane \cite{azcarra00} have
investigated models with only fermionic dynamical variables. Azc\'arraga
{\it et al.} generalises the use of totaly antisymmetric tensors of third
rank in the definition of Killing-Yano tensors and in the construction of
the supercharges of hidden supersymmetries that are at most third in
fermionic variables \cite{azcarra01}.

The SUSY QM formulation has been applied for scattering states (continuum
eigenvalue) in non-relativistic quantum mechanics \cite{Flendey90,chuan90}.
However, a radically different theory for SUSY was recently putted forward,
which is concerned with collision problems in SUSY QM by Shimbori-Kobayashi
\cite{Koba01}.

Zhang {et al.} have considered interesting applications of a semi-unitary
formulation in SUSY QM \cite{zhang00}. Indded, in the papers of Ref.
\cite{zhang00}  a semi-unitary framework of SUSY QM was developed. This
framework works well for multi-dimensional system. Besides Hamiltonian, it
can simultaneously obtain superpartner of the angular momentum and other
observables, though they are not the generators of the superalgebra in SUSY
QM.

Recently, Mamedov {\it et al.} heve applied  SUSY QM  for the case of a
Dirac particle moving in a constant chromomagnetic field \cite{zhang01}.

The spectral resolution for the P\"oschl-Teller potential I has
been studied as shape-invariant potentials and their potential
algebras \cite{Sukhat01}. For this problem we consider as complete
spectral resolution the application of SUSY QM via Hamiltonian
Hierarchy associated to the partner potential respective \cite{pc01r}.

Rencently the group theoretical treatment of SUSY QM has also been
investigated by Fernandez {\it et al.} \cite{fernandez01}. The SUSY
techniques has been applied to periodic potentials  by Dunne-Feinberg
\cite{Dun}, Sukhatme-khare \cite{sukha-khare99} and by Fernandez {\it et
al.} \cite{fernandez00}. Rencently, the complex potentials with the
so-colled PT symmetry in quantum mechanics \cite{znojil00}  has also been
investigated via SUSY QM \cite{Kau01}.

This present work is organized in the following way. In Sec. II we start by
summarizing the essential features of the formulation of one dimensional
supersymmetric quantum mechanics. In Sec. III the factorization of the
unidimensional Schr\"odinger equation and a SUSY Hamiltonian hierarchy
considered by Andrionov {\it et al.} \cite{ABI} and Sukumar \cite{Suku85}
is presented. We consider in Sec. IV the close connection  for SUSY method
as an operator technique for spectral resolution of shape-invariant
potentials. In Sec. V we present our own application of the SUSY
hierarchical prescription for the first P\"oschl-Teller potential. It is
known that the SUSY algebraic method of resolution spectral via property of
shape invariant which permits to work are unbroken SUSY. While the case of
PTPI with broken SUSY in \cite{Sukhat93,Sukhat01} has after suitable
mapping procedures that becomes a new potential with unbroken SUSY, here,
we show that the SUSY hierarchy method \cite{Suku85} can work for both
cases. In Section VI, we present a new scenery on the SUSY when it is
applied for a neutron in interaction with a static magnetic field of a
straight current carrying wire, which is described by two-component wave
functions \cite{VGM,RVV01}.

Section VII contains the concluding remarks.

\section{ N=2 SUSY IN CLASSICAL MECHANICS \protect\\}

Recently, we present a review work on Supersymmetric Classical Mechanics in
the context of a Lagrangian formalism, with $N=1-$supersymmetry.
We have shown
that the $N=1$ SUSY does not allow the introduction of a
potential energy term
depending on a single commuting supercoordinate, $\phi (t;\Theta )$
\cite{RWI}.

In the construction of a SUSY theory with $N>1$, referred to as extended
SUSY, for each space commuting coordinate, representing the degrees of
freedom of the system, we associate one anticommuting variable, which are
known that Grassmannian variables. However, we consider only the $N=2$ SUSY
for a non-relativistic point particle, which is described by the
introduction of two real Grassmannian variables $\Theta_1$ and $\Theta_2$,
in the configuration space, but all the dynamics are putted in the time $t$
\cite{Salo82,junker95,Barce86,Barce97,junker94,CF,Lahi}.

SUSY in classical mechanics is generated by a translation transformation
in the superspace, viz.,

\begin{equation}
\label{GE1}
\Theta_{1} \rightarrow  \Theta^{\prime}_{1} = \Theta _{1}+ \epsilon_{1},
\quad \Theta_{2} \rightarrow  \Theta^{\prime}_{2}= \Theta_{2}+ \epsilon_{2},
\quad
t \rightarrow t^{\prime}= t + i\epsilon_{1}\Theta_{1}+
i\epsilon_{2}\Theta_{2},
\end{equation}
whose are implemented for maintain the line element invariant

\begin{equation}
\label{GE2}
dt-i\Theta _{1}d\Theta _{1}- i\Theta _{2}d\Theta _{2}=\hbox{ invariant},
(\hbox{Jacobian}=1),
\end{equation}
where $\Theta_1, \Theta_2$ and $\epsilon _{1}$ and $\epsilon _{2}$
are real Grassmannian paramenters. We  insert the $i=\sqrt{-1}$ in
(\ref{GE1}) and (\ref{GE2}) to
obtain the real character of time.

The real Grassmannian variables satisfy the
following algebra:

\begin{equation}
\label{GE3}
[\Theta_i, \Theta_j]_+ = \Theta_i\Theta_j + \Theta_j\Theta_i = 0
\Rightarrow  (\Theta_1)^2 = 0 = (\Theta_2)^2.
\end{equation}
They also satisfy the Berezin integral rule \cite{Bere}

\begin{eqnarray}
\label{GE4}
\int d\Theta_i\Theta_j = \delta_{ij}
\Rightarrow\sum^2_{i=1}\int d\Theta_i\Theta_i  = 2,
\quad \int d\Theta_i  = 0
= \partial_{\Theta_i} 1, \quad \int d\Theta_i\Theta_j  = \delta_{ij}
=\partial_{\Theta_i}\Theta_j,
\end{eqnarray}
where
$\partial_{\Theta_i}=\frac{\partial}{\partial\Theta_1}$
so that

\begin{equation}
\label{G5a}
[\partial _{\Theta_i}, \Theta_j ]_+ =
\partial _{\Theta_i} \Theta_j +  \Theta_j \partial _{\Theta_i}=
\delta_{ij}, \quad \partial _{\Theta_i}(\Theta_k\Theta_j)=
\delta_{ik}\Theta_j-\delta_{ij}\Theta_k,
\end{equation}
with  $i=j \Rightarrow
\delta_{ii}=1;$ and if $i\neq j \Rightarrow \delta_{ij}=0,
(i,j=1,2).$

Now, we need to define the derivative rule with respect to one
Grassmannian variable. Here, we use the right derivative rule i.e.
considering $f(\Theta_1,\Theta_2)$ a function of two anticommuting variables,
the right derivative rule is the following:

\begin{eqnarray}
f(\Theta_{\alpha})&=&f_0 +\sum_{\alpha =1}^{2}
f_{\alpha}\Theta_{\alpha}+f_{3}\Theta_{1}\Theta_{2}\nonumber \\
\delta f &=&\sum_{\alpha =1}^{2}\frac{\partial f}{\partial\Theta_{\alpha}}
\delta\Theta_{\alpha}.
\label{GE5}
\end{eqnarray}
where $\delta \Theta_1$ and $\delta \Theta_2$ appear on the right side of
the partial derivatives.

Defining  $\Theta$ and
$\bar{\Theta}$ (Hermitian conjugate of $\Theta$) in terms of
 $\Theta _{i} (i=1, 2)$ and
Grassmannian parameters  $\epsilon_i$,
\begin{eqnarray}
\label{GE6}
\Theta  &=& {1\over \sqrt{2}}(\Theta _{1}- i\Theta _{2}),\nonumber \\
\bar{\Theta } &=& {1\over \sqrt{2}}(\Theta _{1}+ i\Theta _{2}),\nonumber \\
\epsilon  &=& {1\over \sqrt{2}}(\epsilon _{1}- i\epsilon _{2}), \nonumber \\
\bar{\epsilon } &=& {1\over \sqrt{2}}(\epsilon _{1}+ i\epsilon _{2}),
\end{eqnarray}
the supertranslations become:

\begin{equation}
\label{GE7}
\Theta  \rightarrow  \Theta^{\prime}= \Theta  + \epsilon  ,\qquad \bar{\Theta}
\rightarrow \bar{\Theta }^{\prime}= \bar{\Theta } + \bar{\epsilon } ,\qquad
t \rightarrow  t^{\prime}= t - i(\bar{\Theta }\epsilon
-\bar{\epsilon}\Theta).
\end{equation}
In this case, we obtain

\begin{equation}
\label{GE7a}
[\partial_{\Theta}, \Theta ]_+ = 1, \quad
[\partial_{\bar\Theta}, \bar{\Theta} ]_+ =  1, \quad \Theta^2=0.
\end{equation}

The Taylor expansion for the real scalar supercoordinate is given by

\begin{equation}
\label{GE8}
\phi (t;\Theta ,\bar{\Theta }) = q(t) + i\bar{\Theta }\psi (t) + i\Theta
\bar{\psi }(t) + \Theta \bar{\Theta }A(t),
\end{equation}
which under infinitesimal SUSY transformation law provides

\begin{eqnarray}
\delta \phi &=& \phi (t^{\prime};\Theta^{\prime},{\bar\Theta}^{\prime})-
\phi (t;\Theta ,\bar{\Theta })\nonumber\\
{}&=& \partial_t \phi\delta t + \partial_{\Theta} \phi\delta\Theta +
\partial_{\bar\Theta}\phi\delta\bar\Theta \nonumber\\
{}&=& (\bar\epsilon Q + \bar Q\epsilon )\phi
\label{E8a},
\end{eqnarray}
where $\partial_t = \frac{\partial}{\partial t}$ and the two
SUSY generators

\begin{equation}
\label{E8b}
Q \equiv \partial_{\bar\Theta}-i\Theta {\partial_t}, \quad
\bar Q \equiv -\partial_{\Theta}+i\bar{\Theta} {\partial_t}.
\end{equation}
Note that the supercharge $\bar Q$ is not the hermitian conjugate
of the supercharge $Q.$
In terms of $(q(t); A)$ bosonic (even) components
 and $(\psi (t), \bar{\psi }(t))$ fermionic (odd) components we get:

\begin{equation}
\label{GE9}
\delta q(t) = i\{\epsilon \bar{\psi }(t) + \bar{\epsilon }\psi (t)\},
\qquad \delta A = \epsilon \dot{\bar\psi}(t) - \bar{\epsilon} \dot{\psi}(t)
= {d\over dt}\{\epsilon \bar{\psi } - \bar{\epsilon }\psi ),
\end{equation}

\begin{equation}
\label{GE10}
\delta \psi (t) = -\epsilon \{\dot{q}(t) - iA\},\qquad
\delta \bar{\psi }(t) = -\bar{\epsilon }\{\dot{q}(t) + iA\}.
\end{equation}
Therefore making a variation in the even components we
obtain the odd components and vice-versa i.e. SUSY mixes the even and odd
coordinates.

A super-action for the superpoint particle
with N=2 SUSY can be written as the following tripe
integral\footnote{In this section about
supersymmetry we use the unit system in which $m=1=\omega$, where $m$ is
the particle mass and $\omega$ is the angular frequency.}

\begin{equation}
\label{GE11}
S[\phi]= \int\int\int dt d\bar{\Theta} d\Theta
\{{1\over 2}(D\phi )(\bar{D}\phi ) -
U(\phi )\},\qquad \bar D\equiv \partial_{\Theta }
+i\bar{\Theta} \partial_{t},
\end{equation}
where $D$ is the covariant derivative $(D=-\partial_{\bar{\Theta }} -
i\Theta \partial_{t}),\, \bar{\partial}_{\bar\Theta}=-\partial_{\Theta}$ and
$\partial_\Theta = \frac{\partial}{\partial \Theta}$,
built so that $[D, Q]_+ = 0 = [\bar{D}, \bar{Q}]_+$ and $U(\phi )$
is a polynomial function of the supercoordinate.

The covariant derivatives of the supercoordinate
$\phi=\phi(\Theta, \bar{\Theta }; t)$ become

\begin{eqnarray}
\bar D\phi &=& (\partial_{\Theta }+i\bar{\Theta} \partial_{t}) \phi =
-i \bar \psi - \bar \Theta A + i \bar \Theta \partial_t q +
\Theta \bar \Theta\dot{\bar \psi}, \nonumber \\
D\phi &=& (-\partial_{\bar{\Theta }} - i\Theta \partial_{t}) \phi =
i \psi - \Theta A - i\Theta \dot q + \Theta \bar \Theta \dot{\psi} \nonumber \\
(D\phi) (\bar D\phi) &=&
\psi\bar\psi -\bar\Theta(\psi\dot{q}-iA\psi)+
 \Theta(iA\bar\psi +\bar\psi\dot{q})\nonumber \\
&+&\Theta\bar\Theta\left(\dot{q}^{2} + A^{2}
+ i\psi \dot{\bar\psi} +i\dot{\psi }\bar{\psi } \right).
\end{eqnarray}

Expanding in series of Taylor the $U(\phi )$ superpotential and
maintaining
 $ \Theta \bar{\Theta }$ we obtain:

\begin{eqnarray}
\label{GE12}
U(\phi ) &=& \phi U^{\prime}(\phi ) + \frac{\phi^2}{2} U^{\prime\prime} (\phi)
+\cdots \nonumber \\
&=& A \Theta \bar{\Theta }U^{\prime}(\phi ) +
\frac 12 \psi \bar{\psi } \bar{\Theta }\Theta
U^{\prime\prime} + \frac 12 \bar{\psi } \psi \Theta \bar{\Theta }
U^{\prime\prime} + \cdots \nonumber \\
&=& \Theta\bar{\Theta }\{ A U^{\prime} + \bar{\psi } \psi U^{\prime\prime} \}
 + \cdots ,
\end{eqnarray}
where the derivatives  $(U^{\prime}$ and $U^{\prime\prime})$ are such that
$\Theta =0=\bar\Theta $, whose are functions only the $q(t)$ even coordinate.
After the integrations on Grassmannian variables the super-action becomes

\begin{equation}
\label{GE13}
S[q;\psi, \bar\psi] =
{1\over 2}\int\left\{ \dot{q}^{2} + A^{2} - i\dot{\psi }\bar{\psi }
+ i\psi \dot{\bar \psi} -
2AU^{\prime} (q) - 2\bar{\psi }\psi
U^{\prime\prime} (q)\right\}dt \equiv \int Ldt.
\end{equation}
Using the Euler-Lagrange equation to $A$, we obtain:

\begin{equation}
\label{GE14}
\frac{d}{dt} \frac{\partial L}{\partial\partial_t A } -
{\partial L\over \partial A}
= A - U^{\prime}(q) = 0 \Rightarrow A = U^{\prime}(q).
\end{equation}
Substituting  Eq. (\ref{GE14}) in Eq. (\ref{GE13}), we then get the following
Lagrangian for the superpoint particle:

\begin{equation}
\label{GE16}
L = {1\over 2}\left\{\dot{q}^2 -i(\dot\psi\bar{\psi }+\psi\dot{\bar{\psi}})
-2\left(U^{\prime}(q)\right)^{2} -
2U^{\prime\prime}(q)\bar{\psi }\psi\right\},
\end{equation}
where the first term is the kinetic energy associated with the even
coordinate in which the mass of the particle is unity. The second term in
bracket is a kinetic energy piece associated with the odd coordinate
(particle's Grassmannian degree of freedom) dictated by SUSY and is new for
a particle with a potential energy. The Lagrangian is not invariant because
its variation result in a total derivative and consequently is not zero,
however, the super-action is invariant, $\delta S=0,$ which can be obtained
from $D\mid_{\Theta =0}=-Q\mid_{\Theta =0}$ and $\bar D\mid_{\bar\Theta
=0}=-\bar Q\mid_{\bar\Theta =0}.$

The canonical Hamiltonian for the $N=2$ SUSY is given by:

\begin{equation}
\label{GE17}
H_{c}=\dot{q}{\partial L\over \partial \dot{q}} +
{\partial L\over\partial(\partial_t\psi)} \dot\psi +
{\partial L\over \partial
(\partial_t{\bar\psi})} \dot{\bar\psi}  - L = {1\over 2}\left\{p^{2}+
\right(U^{\prime}(q)\left)^{2}+
U^{\prime\prime}(q) [\bar{\psi }, \psi]_{-}\right\},
\end{equation}
which  provides a mixed potential term. Putting $U^{\prime}(x)=-\omega x$
Eq. (\ref{GE13}) describes the super-action for the  superymmetric
oscillator, where $\omega$ is the angular frequency.

\subsection{CANONICAL QUATIZATION IN SUPERSPACE}

The supersymmetry in quantum mechanics, first formulated by Witten \cite{W},
can be deduced via first canonical quantization or Dirac quantization
of above SUSY Hamiltonian which inherently contain constraints.
The first work on the constraint systems without SUSY was
implemented by Dirac in 1950.
The nature of such a constraint is different from the
one encountered in ordinary classical mechanics.

Salomonson {\it et al.} \cite{Salo82}, F. Cooper {\it et al.}, Ravndal
\cite{CF} do not consider such  constraints. However, they have maked an
adequate choice for the fermionic operator representations corresponding to
the odd coordinates $\bar{\psi}$ and  $\psi$. The question of the
constraints in SUSY classical mechanics model have been implemented via
Dirac method  by Barcelos-Neto and Das \cite{Barce86,Barce97}, and by
Junker \cite{junker94}. According the Dirac method the Poisson brackets
$\{A, B\}$ must be substituted by the modified Posion bracket (called Dirac
brackets) $\{A, B\}_D$, which between two dynamic variables $A$ and $B$ is
given by:

\begin{equation}
\label{GE18}
\{A, B\}_{D}= \{A, B\} - \{A, \Gamma _{i}\}C^{-1}_{ij}\{\Gamma _{j},B\}
\end{equation}
where $\Gamma_{i}$  are the second-class constraints. These constraints
define the $C$ matrix

\begin{equation}
\label{GE19}
C_{ij} \simeq  \{\Gamma _{i}, \Gamma _{j}\},
\end{equation}
which Dirac show to be antisymmetric and  nonsingular. The
fundamental canonical Dirac brackets associated with even and odd
coordinates become:

\begin{equation}
\label{GE20}
\{q, \dot{q} \}_D = 1, \qquad \{\psi , \bar{\psi }\}_{D}= i
\qquad\hbox{and}\qquad
\{A, \dot{q}\}_D = {\partial^{2}U(q)\over \partial q^2 }.
\end{equation}
All Dirac brackets  vanish.
It is worth stress that we use the right derivative rule
while Barcelos-Neto and Das in the Ref. \cite{Barce86} have used the
left derivative rule
for the odd coordinates.
Hence unlike of second  Eq. (\ref{GE20}), for odd coordinate
there appears the negative sign in
the corresponding Dirac brackets, i.e., $\{\psi , \bar{\psi }\}_{D}=-i$.

Now in order to implement the first canonical quantization so that
according with the spin-statistic theorem the commutation $[A,
B]_-\equiv AB-BA$ and anti-commutation $[A, B]_+\equiv AB+BA$
relations of quantum mechanics are given by

\begin{eqnarray}
\label{EG21}
\{q, \dot{q}\}_D &=& 1 \rightarrow
{1\over i}[\hat{q} ,\dot{\hat q}]_{-}= 1
\quad \Rightarrow [\hat{q} ,\dot{\hat q}]_{-}=
\hat{q}\dot{\hat q}-\dot{\hat q}\hat{q}=i, \nonumber \\
\{\psi , \bar{\psi }\}_{D}&=&i \rightarrow
\frac 1{-i}[\hat{\psi}, \hat{\bar\psi}]_{+}= i \Rightarrow
[\hat{\psi}, \hat{\bar\psi}]_{+}=\hat{\psi}\hat{\bar\psi}+
\hat{\bar\psi}\hat{\psi}=1.
\end{eqnarray}

Now we will consider the effect of the constraints on the
canonical Hamiltonian in the quantized version. The fundamental
representation of the odd coordinates, in $D=1=(0+1)$ is given by:

\begin{eqnarray}
\label{Eb}
\hat{\psi} &&= \sigma_{+} = \frac{1}{2}(\sigma_{1} +
i\sigma_{2})=\pmatrix{0&0\cr 1&0}\equiv b^+\nonumber\\
\hat{\bar\psi} &&=\sigma_{-} = \frac{1}{2}(\sigma_{1} -
i\sigma_{2})=
\pmatrix{0&1\cr 0&0}\equiv b^-\nonumber\\
{}&&[\hat{\psi}, \hat{\bar{\psi}}]_{+} = {\mbox\large{1}}_{2\times 2},\qquad
[\hat{\bar{\psi}},\hat{\psi}]_{-} = \sigma_{3},
\end{eqnarray}
where $\sigma _{3}$ is the Pauli diagonal matrix,  $\sigma _{1}$ and
 $\sigma _{2}$ are off-diagonal Pauli matrices. On the other
hand, in coordinate representation, it is well known that the
position and momentum operators satisfy the canonical commutation
relation $([\hat x, \hat{p}_x]_-=i)$ with the following
representations:

\begin{equation}
\label{GE24}
\hat x\equiv\hat{q}(t) = x(t), \qquad
\hat{p}_x = m\dot {x}(t) = -i\hbar{d\over dx}
=-i{d\over dx}, \quad \hbar=1.
\end{equation}

In next section we present the various aspects of the SUSY QM and the
connection between Dirac quantization of the SUSY classical mechanics
and the Witten's model of SUSY QM.

\section{The Formulation of SUSY QM}
\label{sec:level1}

The graded Lie algebra satisfied by the odd SUSY charge
 operators $Q_{i}
(i= 1,2,\ldots, N)$ and the even SUSY Hamiltonian $H$ is given by following
 anti-commutation and commutation relations:

\begin{mathletters}
\label{generallabel Q}
\begin{equation}
[Q_{i}, Q_{j}]_{+} = 2 \delta_{ij} H,  \quad(i, j = 1, 2, \ldots, N),
\label{mlett:c}
\end{equation}
\begin{equation}
 [Q_{i}, H]_{-}= 0.
  \label{mlett:d}
\end{equation}
\end{mathletters}
In  these equations, $H$ and $Q_{i}$ are functions of a number of bosonic
and fermionic lowering and raising operators respectively denoted by
$a_{i}, a_{i}^{\dagger} (i = 1, 2, \ldots, N_{b})$ and $b_{i},
b_{i}^{\dagger}(i = 1, 2, \ldots, N_{f})$, that obey the canonical
(anti-)commutation relations:

\begin{mathletters}
\label{generallabel ab}
\begin{equation}
[a_{i}, a^{\dagger}_{j}]_{-} = \delta_{ij},
\label{mlett:a}
\end{equation}
\begin{equation}
[b_{i}, b^{\dagger}_{j}]_{+} = \delta_{ij},
\label{mlett:b}
\end{equation}
\end{mathletters}
all other (anti-)commutators vanish and the bosonic operators always
commute with the fermionic ones.

If we call the generators with these properties "even" and
"odd", respectively, then the SUSY algebra has the general
structure

$$
[even, even]_-=even
$$
$$
[odd, odd]_+=even
$$
$$
[even, odd]_-=odd
$$
which is called a graded Lie algebra or Lie superalgebra by mathematicians.
The case of interest for us is the one with $N_b = N_f = 1$ so that $N =
N_{b} + N_{f} = 2$, which corresponds to the description of the motion of a
spin $\frac{1}{2}$ particle on the real line \cite{W}.

Furthermore, if we
define the mutually adjoint non-Hermitian charge operators

\begin{equation}
\label{E1.3}
Q_{\pm} = \frac{1}{\sqrt{2}} (Q_{1} \pm iQ_{2}),
\end{equation}
in terms of which the Quantum Mechanical SUSY algebraic relations,
 get recast respectively into the following equivalent forms:

\begin{equation}
\label{SA}
Q^2_{+} = Q^2_{-} = 0, \qquad [Q_{+}, Q_{-}]_{+} = H
\end{equation}

\begin{equation}
\label{SC}
[Q_{\pm}, H]_{-} = 0.
\end{equation}
In (\ref{SA})), the nilpotent SUSY charge operators $Q_{\pm}$ and
SUSY Hamiltonian $H$ are now functions of $a^-, a^{+}$ and
$b^-, b^{+}.$ Just as $[Q_i, H]=0$ is a trivial consequence of $[Q_i,Q_j]=
\delta_{ij}H,$ so also
(\ref{SC}) is a direct consequence of (\ref{SA}) and expresses the
invariance of $H$ under SUSY transformations.

We illustrate the same below with the model example of a simple
SUSY harmonic oscillator (Ravndal\cite{CF} and Gendenshtein
\cite{GR}). For the usual bosonic oscillator with the
Hamiltonian\footnote{ NOTATION: Throughout this section, we use
the systems of units such that $c = \hbar = m = 1.$}

\begin{equation}
\label{E5}
H_{b} = \frac{1}{2}\left(p^2_{x} + \omega^2_{b}x^2 \right)
= \frac{\omega_b}{2} [a^{+}, a^{-}]_{+}
= \omega_b \left(N_b + \frac{1}{2}\right), \quad N_b = a^{+} a^{-},
\end{equation}
\begin{equation}
\label{a+-} a^{\pm} = \frac{1}{\sqrt{2\omega_b}} \left(\pm ip_{x} -
\omega_b x \right) = \left(a^{\mp}\right)^\dagger,
\end{equation}
\begin{equation}
\label{E7}
[a^-, a^+]_{-} = 1, \quad [H_b, a^{\pm}]_{-} = \pm \omega_b a^{\pm},
\end{equation}
one obtains the energy eigenvalues

\begin{equation}
\label{E8}
E_b = \omega_b \left(n_{b} + \frac{1}{2}\right), \quad = 0,1,2, \ldots,
\end{equation}
where $n_b$ are the eigenvalues of the number operator indicated here
also by $N_b$.

For the corresponding fermionic harmonic oscillator with the Hamiltonian

\begin{equation}
\label{E9}
H_f = \frac{\omega_{f}}{2}[b^+, b^-]_{-}
= \omega_f \left(N_f - \frac{1}{2}\right),\quad
N_f = b^+ b^-, \quad (b^+)^{\dagger} = b^-,
\end{equation}

\begin{equation}
\label{E10}
[b^-, b^+]_{+} = 1 \quad, (b^-)^{2} = 0 =(b^+)^{2} \quad, [H_f, b^{\pm}]_{-}
=\pm \omega_f b^{\pm},
\end{equation}
we obtain the fermionic energy eigenvalues

\begin{equation}
\label{E11}
E_f = \omega_f \left(\eta_f - \frac{1}{2}\right) \quad, \eta_f = 0,1,
\end{equation}
where the eigenvalues $\eta_f = 0,1$ of the fermionic number operator $N_f$
follow from $N_f{^2} = N_f$.

Considering now the Hamiltonian for the combined system of a bosonic and a
fermionic oscillator with $\omega_b = \omega_f = \omega$, we get:
\begin{equation}
\label{E12}
H = H_b + H_f = \omega\left(N_b + \frac{1}{2} + N_f - \frac{1}{2}\right)
= \omega (N_b + N_f)
\end{equation}
and the energy eigenvalues $E$ of this system are given by the
sum $E_b + E_f$, i.e., by

\begin{equation}
\label{E13}
E = \omega(n_b + n_f) = \omega n, \qquad (n_f = 0, 1; \quad n_b = 0, 1, 2,
\ldots; \quad n = 0, 1, 2, \ldots).
\end{equation}
Thus the ground state energy $E^{(0)} = 0$ in (\ref{E13}) corresponds to the
only non-degenerate case with $n_b = n_f = 0$, while all the excited state
energies $E^{(n)} (n\geq 1)$ are doubly degenerate with $(n_b, n_f) = (n,
0)$ or $(n-1, 1)$, leading to the same energy $E^{(n)} = n\omega$ for
$n\geq 1.$

The extra symmetry of the Hamiltonian (\ref{E12}) that leads to the above of
double degeneracy (except for the singlet ground state) is in fact a
supersymmetry, i.e., one associated with the simultaneous destruction of
one bosonic quantum $n_b \rightarrow n_b-1$ and creation of one fermionic
quantum $n_f \rightarrow n_f+1$ or vice-versa, with the corresponding
symmetry generators behaving like $a^- b^+$ and $a^+ b^-.$ In fact,
defining,

\begin{equation}
\label{E14a}
Q_{+} = \sqrt {\omega} a^+ b^-, \qquad Q_{-} = (Q_{+})^\dagger =
\sqrt {\omega} a^- b^+,
\end{equation}
it can be directly verified that these charge operators satisfy the SUSY
algebra given by Eqs. (\ref{SA}) and (\ref{SC}).

Representing the fermionic operators by Pauli matrices
as given by Eq. (\ref{Eb}), it follows that

\begin{equation}
\label{ENF} N_f = b^{+}b^{-} = \sigma_{-}\sigma_{+} = \frac{1}{2}(1
-\sigma_{3}),
\end{equation}
so that the Hamiltonian (12) for the SUSY harmonic oscillator takes the
following form:

\begin{equation}
\label{OSSa}
H = \frac{1}{2}p^{2}_x + \frac{1}{2}\omega^{2} x^{2} - \frac{1}{2}
\sigma_{3}\omega,
\end{equation}
which resembles the one for a spin $\frac{1}{2}$ one dimensional harmonic
oscillator subjected to a constant magnetic field. Explicitly,

\begin{eqnarray}
\label{OSSb}
H &&= \left(
\begin{array}{cc}
\frac{1}{2}p^{2}_x + \frac{1}{2}\omega^{2} x^{2} - \frac{1}{2}\omega & 0 \\
0 & \frac{1}{2}p^{2}_x + \frac{1}{2}\omega^{2} x^{2} + \frac{1}{2}\omega
\end{array}\right)
\nonumber\\
{}&&= \left(
\begin{array}{cc}
\omega a^+a^- & 0 \\
0 & \omega a^- a^+
\end{array}\right)= \left(
\begin{array}{cc}
H_- & 0 \\
0 & H_+
\end{array}\right)
\end{eqnarray}
where, from Eqs. (\ref{Eb}) and (\ref{E14a}), we get

\begin{equation}
\label{E22}
Q_{+} = \sqrt{\omega}
\left(
\begin{array}{cc}
0 & a^{+} \\
0 & 0
\end{array}\right),\qquad
Q_{-} = \sqrt {\omega}
\left(
\begin{array}{cc}
0 & 0 \\
a^{-} & 0
\end{array}\right).
\end{equation}
The eigenstates of $N_f$ with the fermion number $n_f = 0$ is called
bosonic states and is given by

\begin{equation}
\label{E23a}
\chi^{-} = \chi^{\uparrow}
= \left(
\begin{array}{cc}
1 \\
0
\end{array}\right).
\end{equation}
Similarly, the eigenstates of $N_f$ with the fermion number $n_f = 1$
is called fermionic state and is given by

\begin{equation}
\label{E23b}
\chi^{+} = \chi^{\downarrow}
= \left(
\begin{array}{cc}
0 \\
1
\end{array}\right).
\end{equation}
The subscripts $-(+)$ in $\chi_-(\chi_+)$ qualify their
non-trivial association with $H_-(H_+)$ of $H$ in (\ref{OSSb}).
Accordingly, $H_-$ in (\ref{OSSb}) is said to refer to the bosonic
sector of the SUSY Hamiltonian $H$ while $H_+$, the fermionic
sector of $H$. (Of course this qualification is only conventional
as it depends on the mapping adopted in (\ref{Eb}) of $b^\mp$ onto
$\sigma^\pm$, as the reverse mapping is easily seen to reverse the
above mentioned qualification.)

\subsection{WITTEN'S QUANTIZATION WITH SUSY}

Witten's model \cite{W} of the one dimensional SUSY quantum system is a
generalization of the above construction of a SUSY simple harmonic oscillator
with $\sqrt{\omega} a^- \rightarrow A^-$ and $\sqrt{\omega} a^+ \rightarrow
A^+$, where

\begin{equation}
\label{E24}
A^{\mp} = \frac{1}{\sqrt{2}} (\mp ip_{x} - W(x)) = (A^{\pm})^{\dagger},
\end{equation}
where, $W = W(x)$, called the superpotential, is an arbitrary function of
the position coordinate. The position $x$ and its canonically conjugate
momentum $p_x = -i\frac{d}{dx}$ are related to $a^-$ and $a^+$ by
(\ref{a+-}), but with $\omega_b = 1$:

\begin{equation}
\label{E25}
a^{\mp} = \frac{1}{\sqrt{2}} (\mp ip_{x} - x) = (a^{\pm})^\dagger.
\end{equation}
The mutually adjoint non-Hermitian supercharge operators for
Witten's model \cite{W,Fred} are given by

\begin{equation}
\label{E26} Q_{+} = A^+\sigma_-=\left(
\begin{array}{cc}
0 & {A^+} \\
0 & 0
\end{array}\right),\quad
Q_{-} = A^-\sigma_+\left(
\begin{array}{cc}
0 & 0 \\
{A^-} & 0
\end{array}\right),
\end{equation}
so that the SUSY Hamiltonian $H$ takes the form

\begin{eqnarray}
\label{HS}
H = [Q_+,Q_- ]_+
&=& \frac{1}{2} \left(p^{2}_x + W^{2}(x) -\sigma _{3} \frac{d}{dx} W(x)\right)
\nonumber \\
&=& \left(
\begin{array}{cc}
H_{-} & 0 \\
0 & H_{+}
\end{array}\right)
= \left(
\begin{array}{cc}
A^{+}A^{-} & 0 \\
0 & A^{-}A^{+}
\end{array}\right)
\end{eqnarray}
where $\sigma_3$ is the Pauli diagonal matrix and, explicitly,

\begin{eqnarray}
\label{E28a}
H_- &&= A^+ A^-
= \frac{1}{2} \left(p^{2}_x + W^{2}(x) - \frac{d}{dx}W(x)\right)\nonumber\\
H_+ &&= A^- A^+
= \frac{1}{2} \left(p^{2}_x + W^{2}(x) + \frac{d}{dx}W(x)\right).
\end{eqnarray}

In this stage we present the connection between the Dirac quantization
and above SUSY Hamiltonian. Indeed, from Eq. (\ref{Eb}) and (\ref{GE17}),
and defining

\begin{equation}
\label{GE25}
W(x) \equiv  U^{\prime} (x) \equiv  {dU\over dx},
\end{equation}
the SUSY Hamiltonian given by Eq. (\ref{HS}) is reobtained.

Note that for the choice of $W(x) =\omega x$ one reobtains the
unidimensional SUSY oscillator (\ref{OSSa}) and (\ref{OSSb}) for which

\begin{eqnarray}
\label{E29}
A^- =a^-&&= \frac{1}{\sqrt {2}} \left(-\frac{d}{dx} - \omega x
\right) = \psi^{(0)}_- \left(-\frac{1}{\sqrt {2}} \frac{d}{dx} \right)
\frac{1}{\psi^{(0)}_-}\nonumber\\
 A^+ =a^+&&= (A^{-})^{\dagger} =
\frac{1}{\sqrt {2}} \left(\frac{d}{dx} - \omega x \right)
=\frac{1}{\psi^{(0)}_-} \left(\frac{1}{\sqrt {2}} \frac{d}{dx} \right)
\psi^{(0)}_-,
\end{eqnarray}
where

\begin{equation}
\label{E30}
\psi^{(0)}_- \propto exp\left(-\frac{1}{2}\omega x^2\right)
\end{equation}
is the normalizable ground state wave function of the bosonic
sector Hamiltonian $H_-$.

In an analogous manner, for the SUSY Hamiltonian (\ref{HS}), the operators
$A^{\pm}$ of (\ref{E24}) can be written in the form

\begin{eqnarray}
\label{E31a}
A^- &&= \psi^{(0)}_- \left(-\frac{1}{\sqrt {2}} \frac{d}{dx} \right)
\frac{1}{\psi^{(0)}_-}\nonumber\\
{}&&= \frac{1}{\sqrt {2}} \left(-\frac{d}{dx} + \frac{1}{\psi^{(0)}_-}
\frac{d\psi_{-}^{(0)}}{dx} \right)
\end{eqnarray}

\begin{eqnarray}
A^+ &&= (A^-)^{\dagger} = \frac{1}{\psi^{(0)}_-}
\left(\frac{1}{\sqrt {2}}
\frac{d}{dx} \right)\psi^{(0)}_-\nonumber\\
{} &&=\frac{1}{\sqrt {2}}
\left(\frac{d}{dx} +\frac{1}{\psi^{(0)}_-}\frac{d\psi_{-}^{(0)}}{dx} \right),
\end{eqnarray}
where

\begin{equation}
\psi^{(0)}_- \propto exp\left(-\int^{x}W(q)dq\right)
\label{E32a}
\end{equation}
and

\begin{equation}
\psi^{(0)}_+ \propto exp\left(\int^{x}W(q)dq
\right) \Rightarrow  \psi^{(0)}_+ \propto\frac{1}{\psi^{(0)}_-}
\label{E32b}
\end{equation}
are symbolically the ground states of $H_-$ and $H_+$, respectively.
Furthermore, we may readily write the following annihilation conditions
for the operators $A^{\pm}$:

 \begin{equation}
A^-\psi^{(0)}_-=0, \qquad A^+\psi^{(0)}_+=0.
\label{CAA}
\end{equation}

Whatever be the functional form of $W(x)$, we have, by virtue of Eqs.
(\ref{E23a}), (\ref{E23b}), (\ref{E26}), (\ref{E32a}), (\ref{E32b}) and
(\ref{CAA}),

\begin{equation}
\label{E33a}
Q_{-} \psi^{(0)}_{-} \chi_{-}
= 0, \qquad |\phi_{-}> \equiv \psi^{(0)}_{-} \chi_{-} \propto
exp\left(-\int^{x} W(q)dq\right)
\left(
\begin{array}{cc}
1 \\
0
\end{array}\right)
\end{equation}
and

\begin{equation}
\label{E33b}
Q_{+} \psi^{(0)}_{+} \chi_{+}
= 0,\qquad |\phi_{+}> \equiv \psi^{(0)}_{+} \chi_{+} \propto
\hbox{exp}\left(\int^{x}W(q)dq\right)
\left(
\begin{array}{cc}
0 \\
1
\end{array}\right),
\end{equation}
so that the  eigensolution $|\phi_{-}>$ and $|\phi_{+}>$ of
(\ref{E33a}) and (\ref{E33b}) are both annihilated by the SUSY
Hamiltonian (\ref{HS}), with $Q_{-} \chi_{+} = 0$ and $Q_{+}
\chi_{-} = 0$, trivially holding good. If only one of these
eigensolution, $|\phi_{-}>$ or $|\phi_{+}>$, are normalizable, it
then becomes the unique eigenfunction  of the SUSY Hamiltonian
(\ref{HS}) corresponding to the zero energy of the ground state.
In this situation, SUSY is said to be unbroken. In the case when
neither $|\phi_{-}>$, Eq. (\ref{E33a}), nor $|\phi_{+}>$, Eq.
(\ref{E33b}), are normalizable, then no normalizable zero energy
state exists and SUSY is said to be broken. It is readily seen
from (\ref{E33a}) and (\ref{E33b}) that if $W(x) \rightarrow
\infty(-\infty)$, as $x \rightarrow \pm \infty$, then
$|\phi_{-}>(|\phi_{+}>)$ alone is normalizable with unbroken SUSY
while for $W(x) \rightarrow -\infty$ or $+\infty$, for $x
\rightarrow \pm\infty$ neither $|\phi_{-}>$ nor $|\phi_{+}>$ are
normalizable and one has broken SUSY dynamically
\cite{W,A,G,Fred}. In this case, there are no zero energy for the
ground state and so far the spectra to $H_{\pm}$ are identical.

Note from the form of the SUSY Hamiltonian $H$ of (\ref{HS}), that the two
second-order differential equations corresponding to the eigenvalue
equations of $H_{-}$ and $H_{+}$ of Eq. (\ref{E28a}), by themselves
apparently unconnected, are indeed related by SUSY transformations by
$Q_{\pm}$, Eq. (\ref{E26}), on $H$, which operations get translated in
terms of the operators $A^{\pm}$ in $Q_{\pm}$ as discussed below.

\subsubsection{FACTORIZATION OF THE SCHR\"ORDINGER AND A SUSY HAMILTONIAN}
\label{sec:level2}

Considering the case of unbroken SUSY and observing that the SUSY
Hamiltonian (\ref{HS}) is invariant under $x \rightarrow -x$ and
$W(x) \rightarrow -W(x)$
there is no loss of generality involved in assuming that $|\phi_{-}>$
of (\ref{E33a}) is the normalizable ground state wave function of $H$ so that
$\psi^{(0)}_{-}$ is the ground state wave function of $H_{-}.$ Thus, from
(\ref{HS}), (\ref{E28a}), (\ref{E31a}) and (\ref{E33a}), it follows that

\begin{equation}
\label{E34}
H_{-} \psi^{(0)}_{-}
= \frac{1}{2} \left(p^{2}_x + W^{2}(x) - W^{\prime}(x)
\right)\psi^{(0)}_{-}
= A^+ A^- \psi^{(0)}_{-} = 0,
\end{equation}

\begin{equation}
E^{(0)}_ -  = 0, \quad V_{-}(x) = \frac{1}{2}W^2(x)
- \frac{1}{2}W^{\prime} (x), \quad
W^{\prime} (x) = \frac{d}{dx}W(x).
\label{E35}
\end{equation}
Them from (\ref{HS}) and (\ref{E29}),

\begin{equation}
H_{+} = A^- A^+ = A^+ A^- - [ A^+, A^-]_-
= H_{-} -\frac{d^{2}}{dx^{2}} \ell n \psi^{(0)}_{-},
\label{E36}
\end{equation}

\begin{equation}
V_{+}(x) = V_{-}(x) - \frac{d^{2}}{dx^{2}}\ell n \psi^{(0)}_{-}
= \frac{1}{2}W^2 (x) + \frac{1}{2}W^{\prime} (x).
\label{E37}
\end{equation}

From (\ref{E34}) and (\ref{E35}) it is clear that any Schr\"odinger
equation with potential $V_-(x)$, that can support at least one bound state
and for which the ground state wave function $\psi^{(0)}_{-}$ is known, can
be factorized in the form (\ref{E34}) with $V_-$ duly readjusted to give
$E^{(0)}_{-} = 0$ (Andrianov {\it et al.} \cite{ABI} and Sukumar
\cite{Suku85}). Given any such readjusted potential $V_{-}(x)$ of
(\ref{E35}), that supports a finite number, $M$, of bound states, SUSY
enables us to construct the SUSY partner potential $V_{+}(x)$ of
(\ref{E37}). The two Hamiltonians $H_{-}$ and $H_{+}$ of (\ref{HS}),
(\ref{E34}) and (\ref{E36}) are said to be SUSY partner Hamiltonians. Their
spectra and eigenfunctions are simply related because of SUSY invariance of
$H$, i.e., $[Q_{\pm},H]_{-} =0.$

Denoting the eigenfunctions of $H_{-}$ and $H_{+}$ respectively by
$\psi^{(n)}_{-}$ and $\psi^{(n)}_{+}$, the integer
$n= 0,1,2 \ldots$ indicating the number of nodes in the wave function,
we show now that $H_{-}$ and $H_{+}$
possess the same energy spectrum, except that the ground state energy
$E^{(0)}_{-}$ of $V_{-}$ has no corresponding level for $V_{+}$.

Starting with
\begin{equation}
\label{E388}
H_{-}\psi^{(n)}_{-} = E^{(n)}_{-}\psi^{(n)}_{-}
\Longrightarrow  A^+ A^- \psi^{(n)}_{-} =  E^{(n)}_{-}\psi^{(n)}_{-}
\end{equation}
and multiplying (\ref{E388}) from the left by $A^{-}$ we obtain

\begin{equation}
A^- A^+(A^- \psi^{(n)}_{-}) = E^{(n)}_{-} (A^- \psi^{(n)}_{-})
\Rightarrow H_{+}(A^- \psi^{(n)}_{-}) =  E^{(n)}_{-}
(A^- \psi^{(n)}_{-}).
\label{E39}
\end{equation}
Since $A^{-} \psi^{(0)}_{-} = 0$ [see Eq. (\ref{CAA})], comparison
of (\ref{E39}) with

\begin{equation}
H_{+}\psi^{(n)}_{+} = A^- A^+ \psi^{(n)}_{+} =
E^{(n)}_{+}\psi^{(n)}_{+},
\label{E40}
\end{equation}
leads to the immediate mapping:

\begin{equation}
E^{(n)}_{+} = E^{(n+1)}_{+}, \quad\psi^{(n)}_{+}\propto A^{-}
\psi^{(n+1)}_{-}, \quad  n = 0, 1, 2, \ldots. \label{E41}
\end{equation}

Repeating the procedure but starting with (\ref{E40}) and
multiplying the same from the left by $A^{+}$ leads to

\begin{equation}
\label{generallabel }
A^+ A^- (A^+ \psi^{(n)}_{+}) = E^{(n)}_{+} (A^+ \psi^{(n)}_{+}),
\label{E42}
\end{equation}
so that it follows from (\ref{E388}), (\ref{E41}) and (\ref{E42}) that

\begin{equation}
\psi^{(n+1)}_{-}\propto A^{+} \psi^{(n)}_{+}, \quad n = 0, 1, 2, \ldots.
\label{E43}
\end{equation}
The intertwining operator $A^{-}(A^{+})$ converts an eigenfunction of
$H_{-}(H_{+})$
into an eigenfunction of $H_{+}(H_{-})$ with the same energy and
simultaneously destroys (creates) a node of $\psi^{(n+1)}_{-}
\left(\psi^{(n)}_{+} \right).$ These operations just express the content of
the SUSY operations effected by $Q_{+}$ and $Q_{-}$
of (\ref{E26}) connecting the bosonic and fermionic sectors of the
SUSY Hamiltonian (\ref{HS}).

The SUSY analysis presented above in fact enables the generation of a
hierarchy of Hamiltonians with the eigenvalues and the eigenfunctions of the
different members of the hierarchy in a simple manner (Sukumar \cite{Suku85}).
Calling  $H_{-}$ as $H_{1}$ and $H_{+}$ as $H_{2}$, and suitably
changing the subscript qualifications, we have

\begin{equation}
H_1 = A^{+}_1 A^{-}_1 + E^{(0)}_1, \quad A^{(-)}_1 = \psi^{(0)}_{1}
\left(-\frac{1}{\sqrt {2}}\frac{d}{dx}\right) \frac{1}{\psi^{(0)}_{1}}
= (A^{+}_1)^{\dagger}, \quad E^{(0)}_1 = 0,
\label{E44}
\end{equation}
with supersymmetric partner given by

\begin{equation}
H_2 = A^{-}_1 A^{+}_1 + E^{(0)}_1 , \qquad
V_2 (x) = V_1 (x) - \frac{d^{2}}{dx^{2}}\ell n \psi^{(0)}_{1}.
\label{E45}
\end{equation}
The spectra of $H_{1}$ and $H_{2}$ satisfy [see (\ref{E41})]
\begin{equation}
E^{(n)}_2 =  E^{(n+1)}_1, \qquad n = 0, 1, 2, \ldots,
\label{E46}
\end{equation}
with their eigenfunctions related by [see (\ref{E43})]
\begin{equation}
\psi^{(n+1)}_{1} \alpha  A^{+}_1 \psi^{(n)}_2, \qquad n = 0, 1, 2, \ldots .
\label{E47}
\end{equation}

Now factoring $H_{2}$ in terms of its ground state wave
function $\psi^{(0)}_{2}$ we have

\begin{equation}
H_2 = -\frac{1}{2}\frac{d^{2}}{dx^{2}} + V_2 (x) = A^{+}_2 A^{-}_2 +
E^{(0)}_2 ,\quad
A^{-}_2 = \psi^{(0)}_2 \left(-\frac{1}{\sqrt {2}}\frac{d}{dx}\right)
\frac{1}{\psi^{(0)}_{2}},
\label{E48}
\end{equation}
and the SUSY partner of $H_{2}$ is given by

\begin{equation}
H_3 = A^{-}_2 A^{+}_2 + E^{(0)}_2 ,  \quad V_3 (x) = V_2 (x) -
\frac{d^{2}}{dx^{2}}\ell n \psi^{(0)}_{1}. \label{E49}
\end{equation}
The spectra of $H_{2}$ and $H_{3}$ satisfy the condition

\begin{equation}
E^{(n)}_3 =  E^{(n+1)}_2,  \quad n = 0, 1, 2, \ldots, \label{E50}
\end{equation}
with their eigenfunctions related by

\begin{equation}
\psi^{(n+1)}_{2} \alpha  A^{+}_2 \psi^{(n)}_3,  n = 0, 1, 2, \ldots.
\label{E51}
\end{equation}

Repetition of the above procedure for a finite number, $M,$ of bound states
leads to the generation of a hierarchy of Hamiltonians given by

\begin{equation}
H_n = -\frac{1}{2}\frac{d^{2}}{dx^{2}} + V_n (x) = A^{+}_n A^{-}_n + E^{(0)}_n
= A^{-}_{n-1} A^{+}_{n-1} + E^{(0)}_{n-1} ,
\label{E52}
\end{equation}
where

\begin{eqnarray}
\label{E53a}
A^{-}_n &&= \psi^{(0)}_n \left(-\frac{1}{\sqrt
{2}}\frac{d}{dx}\right) \frac{1}{\psi^{(0)}_{n}} = \frac{1}{\sqrt
{2}}\left(-\frac{d}{dx}-W_n(x)\right), \nonumber\\
W_n(x)&&=-\frac{d}{dx}\ell n(\psi^{(0)}_{n}),
\quad A^+_n=\left(A^{-}_n\right)^{\dagger},
\end{eqnarray}
and

\begin{eqnarray}
\label{E54}
V_n (x) &&= V_{n-1} (x) - \frac{d^{2}}{dx^{2}}
\ell n (\psi^{(0)}_{n-1}) \nonumber\\
{}&&= V_{1} (x) - \frac{d^{2}}{dx^{2}}
\ell n (\psi^{(0)}_{1} \psi^{(0)}_{2} \ldots \psi^{(0)}_{n-1}), \quad
n=2,3,\ldots, M,
\end{eqnarray}
whose spectra satisfy the conditions

\begin{equation}
E^{n-1}_{1} = E^{n-2}_{2} = \ldots = E^{(0)}_{n}, \qquad n=2,3,\ldots,M,
\label{E55a}
\end{equation}

\begin{equation}
\psi^{n-1}_{1}  \propto  A^{+}_{1} A^{+}_{2} \ldots A^{+}_{n-1} \psi^{(0)}_n.
\label{E55b}
\end{equation}
Note that the nth-member of the hierarchy has the same eigenvalue spectrum
as the first member $H_{1}$ except for the missing of the first $(n-1)$
eigenvalues of $H_{1}.$ The energy eigenvalue of the (n-1)th-excited state
of  $H_{1}$ is degenerate with the ground state of $H_{n}$ and can be
constructed with the use of (\ref{E55b}) that involves the knowledge of
$A_{i}(i=1,2, \ldots, n-1)$ and $\psi^{(0)}_{n}.$

\subsubsection{SUSY METHOD AND SHAPE-INVARIANT POTENTIALS}
\label{sec:level3}

It is particularly simple to apply (\ref{E55b}) for shape-invariant potentials
(Gendenshtein \cite{Gend},
Cooper {\it et al.} \cite{CGK}, Dutt {\it et al.} \cite{Sukhat88} and
in review article \cite{Fred})
as their SUSY partners  are similar in shape and differ only in the
parameters that appear in them. More specifically, if $V_{-}(x;a_{1})$ is
any potential, adjusted to have zero ground state energy $E^{(0)}_{-} = 0,$
its SUSY partner
$V_{+}(x;a_{1})$ must satisfy the requirement

\begin{equation}
V_+ (x;a_1) = V_- (x;a_2) + R(a_2), \qquad a_2 = f(a_1),
\label{E56}
\end{equation}
where $a_{1}$ is a set of parameters, $a_{2}$ a function of the
parameters $a_{1}$ and $R(a_{2})$ is a remainder independent of $x$.
Then, starting with $V_{1} = V_{-}(x;a_{2})$ and
$V_{2} = V_{+}(x;a_{1}) = V_{1}(x;a_{2}) + R(a_{2})$ in (\ref{E56}),
one constructs a hierarchy of Hamiltonians

\begin{equation}
H_n = -\frac{1}{2}\frac{d^{2}}{dx^{2}} + V_{-} (x;a_n) +
\Sigma_{s=2}^{n} R(a_s),
\label{E57}
\end{equation}
where $a_{s} = f^{s}(a_{1})$, i.e., the function $f$
applied $s$ times. In view of Eqs. (\ref{E56}) and (\ref{E57}), we have

\begin{equation}
H_{n+1}
= -\frac{1}{2}\frac{d^{2}}{dx^{2}} + V_{-}(x;a_{n+1}) +
\Sigma_{s=2}^{n+1}R(a_s)
\label{E58a}
\end{equation}

\begin{equation}
= -\frac{1}{2}\frac{d^{2}}{dx^{2}} + V_{+}(x;a_{n}) + \Sigma_{s=2}^{n}R(a_s).
\label{E58b}
\end{equation}
Comparing (\ref{E57}), (\ref{E58a}) and (\ref{E58b}), we immediately note
that $H_{n}$ and
$H_{n+1}$ are SUSY partner Hamiltonians with identical energy spectra
except for the ground state level

\begin{equation}
E^{(0)}_n = \Sigma_{s=2}^{n}R(a_s)
\label{E59}
\end{equation}
of $H_{n}$, which follows from Eq. (\ref{E57}) and the
normalization that for any
$V_{-}(x;a)$ , $E^{(0)}_{-} = 0$. Thus Eqs. (\ref{E55a}) and
(\ref{E55b})
get translate simply, letting $n \rightarrow n+1$, to

\begin{equation}
E^{n}_1 = E^{n-1}_{2} = \ldots = E^{(0)}_{n+1} = \sum_{s=2}^{n+1}R(a_s),
\quad
 n = 1, 2, \ldots
\label{E60a}
\end{equation}
and

\begin{equation}
\psi^{(n)}_{1}\propto A^{+}_1(x;a_1) A^{+}_2(x;a_2) \ldots A^{+}_n(x;a_n)
\psi^{(0)}_{n+1}(x;a_{n+1}).
\label{E60b}
\end{equation}

Equations (\ref{E60a}) and (\ref{E60b}), succinctly express the SUSY
algebraic generalization, for various shape-invariant potentials of
physical interest \cite{Gend,Suku85,Sukhat88}, of the method of
constructing energy eigenfunctions $(\psi^{(n)}_{osc})$ for the usual ID
oscillator problem. Indeed, when $a_1=a_2= \ldots =a_n=a_{n+1},$ we obtain
$\psi^{(n)}_{osc}\propto (a^+)^n \psi^{(0)}_{1}, \quad A^+_n=a^+, \quad
\psi^{(0)}_{osc}=\psi^{(0)}_{n+1}=\psi^{(0)}_1 \propto e^{-\frac{\omega
x^2}{2}},$ where $\omega$ is the angular frequency.

The shape invariance has an underlying algebraic structure and may
be associated with Lie algebra \cite{balan98}.
In next Section  of this work, we present our own application of the Sukumar's
SUSY method outlined above for the first P\"oschl-Teller potential with
unquantized coupling constants,
while in the earlier SUSY algebraic treatment by
Sukumar \cite{Suku85} only the restricted symmetric case of this potential with
quantized coupling constants was considered.

\section{THE FIRST P\"OSCHL-TELLER
POTENTIAL VIA SUSY QM}
\label{sec:level6}

We would like to stress the interesting approaches for the
P\"oschl-Teller I potential. Utilizing the SUSY connection between
the particle in a box with perfectly rigid walls and the symmetric
first P\"oschl-Teller potential, the SUSY hierarchical
prescription (outlined in Section III) was utilized by Sukumar
\cite{Suku85} to solve  the energy spectrum of this potential. The
unsymmetric case has recently been treated algebraically by Barut,
Inomata and Wilson \cite{BIW}. However in these analysis only
quantized values of the coupling constants of the P\"oschl-Teller
potential have been obtained. In the works of Gendenshtein
\cite{Gend} and Dutt {\it et al.} \cite{Sukhat88} treating the
unsymmetric case of this potential with unquantized coupling
constants by the SUSY method for shape-invariant potentials, only
the energy spectrum was obtained but not the excited state wave
functions. Below we present our own application of the Sukumar's
SUSY method obtaining not only the energy spectrum but also the
complete excited state energy eigenfunctions.

 It is well known that usual shape invariance procedure \cite{Gend} is not
applicable for computation energy spectrum of a potential without zero
energy eigenvalue. Recently, an approach was implemented with a two-step
shape invariant in order to connect broken and unbroken SUSY QM potentials
\cite{Sukhat93,Sukhat01}. In this references it is   considered the
P\"oschl-Teller I potential, showing the types of shape invariance it
possesses. In Ref.  \cite{Sukhat01}, the PTPI and the three-dimensional
harmonic oscillator both with broken SUSY have been investigated, for the
first time, in terms of a novel two-step shape invariance approach via a
group theoretic potential algebra approach \cite{balan98}. In the work
present is the first spectral resolution, to our knowledge, via SUSY
hierarchy in order to construct explicitly the energy eigenvalue and
eigenfunctions of the P\"oschl-Teller I potential.

Starting with the first P\"oschl-Teller Hamiltonian \cite{Flu}

\begin{equation}
\label{PT}
H_{PT} = -\frac{1}{2}\frac{d^2}{dx^2} + \frac{1}{2}\alpha^{2}
\left\{\frac{k(k-1)}{sin^{2}\alpha{x}}
+ \frac{\lambda(\lambda-1)}{cos^{2}\alpha{x}}\right\},
\end{equation}
where $0 \leq \alpha{x}\leq\pi/2, k>1, \lambda > 1;
\alpha =\hbox{real constant}.$ The
substitution $\Theta = 2\alpha {x}, 0 \leq \Theta \leq \pi,$ in (\ref{PT})
leads to

\begin{equation}
\label{NPT}
H_{PT} = 2 \alpha^{2}H_1
\end{equation}
where

\begin{eqnarray}
\label{PT1}
H_1 &&= - \frac{d^2}{d\Theta^{2}} + V_1(\Theta)\nonumber\\
V_1(\Theta)&&=\frac{1}{4}
\left[k(k-1)sec^{2}(\Theta/2) +
\lambda(\lambda-1)cossec^{2}(\Theta/2)\right].
\end{eqnarray}

Defining

\begin{equation}
\label{A1}
A_1^{\pm}=\pm\frac{d}{d\Theta}-W_1(\Theta)
\end{equation}
and

\begin{eqnarray}
H_1&&= A^{+}_{1} A^{-}_{1} + E^{(0)}_{1}\nonumber\\
{}&&=-\frac{d^2}{d\Theta^{2}}+W^2_1(\Theta)-W_1^{\prime}(\Theta)+
E^{(0)}_{1}
\end{eqnarray}
where the prime means a first derivative with respect to $\Theta$
variable.
From both above definitions of $H_1$ we obtain the following
non-linear first order differential equation

\begin{equation}
\label{R1}
W_1^2(\theta)- W_1^{\prime}(\Theta)=\frac{1}{4}
\left\{\frac{k(k-1)}{sin^{2}(\Theta/2)} +
\frac{\lambda(\lambda-1)}{cos^{2}(\Theta/2)}\right\}-E^{(0)}_{1},
\end{equation}
which is exactly a Riccati equation.

Let be superpotential Ansatz
\begin{equation}
\label{W1}
W_1(\Theta)=
-\frac k2 cot(\Theta/2) +\frac{\lambda}{2}tan(\Theta/2), \quad
E^{(0)}_{1} = \frac{1}{4} (k+\lambda)^{2}.
\end{equation}
According to Sec. III, the energy eigenfunction associated to the
ground state  of PTI potential becomes

\begin{equation}
\label{GS1}
\psi^{(0)}_{1}=exp\left\{-\int W_1(\theta)d\Theta\right\}\propto
sin^k(\Theta/2)cos^{\lambda}(\Theta/2).
\end{equation}
In this case the first order intertwining operators become

\begin{equation}
\label{A1-}
A^{-}_{1} = -\frac{d}{d\Theta} +
\frac{k}{2}cot(\Theta/2)- \frac{\lambda}{2}tan(\Theta/2) = \psi^{(0)}_1
\left(-\frac{d}{d\Theta}\right)\frac{1}{\psi^{(0)}_1}
\end{equation}
and
\begin{eqnarray}
\label{A1+}
A^{+}_{1} &&= (A^{-}_{1})^\dagger =
\frac{1}{\psi^{(0)}_1} \left(\frac{d}{d\Theta}\right)\psi^{(0)}_{1}\nonumber\\
{}&&= \frac{d}{d\Theta} + \frac{k}{2}cot(\Theta/2)-
\frac{\lambda}{2}tan(\Theta/2).
\end{eqnarray}
In Eqs. (\ref{A1-})) and (\ref{A1+}), $\psi^{(0)}_1$
is the ground state wave function of $H_{1}$.

The SUSY partner of $H_{1}$ is $H_{2}$, given by

\begin{eqnarray}
\label{H12}
H_2 &&= A^{-}_{1} A^{+}_{1} + E^{(0)}_{1} =
H_{1} - [A^+_{1}, A^-_{1}]_-\nonumber\\
 V_2(\Theta) &&= V_{1}(\Theta) -
2\frac{d^2}{d\Theta^2}\ell n\psi^{(0)}_1\nonumber\\
{}&&= V_1(\Theta) - 2\frac{d^2}{d\Theta^2}
\ell n \left[sin^k(\Theta/2) cos^\lambda(\Theta/2)\right]\nonumber\\
{}&&= \frac{1}{4} \left(\frac{k(k+1)}{sin^{2}(\Theta/2)} +
\frac{\lambda(\lambda+1)}{cos^{2}(\Theta/2)}\right).
\end{eqnarray}

Let us now consider a refactorization of $H_2$ in its ground state

\begin{equation}
\label{H2N}
H_2 = A^{+}_{2} A^{-}_{2} + E^{(0)}_2,  \quad
A^{-}_{2} = -\frac{d}{d\Theta} -W_2(\Theta).
\end{equation}
In this case we find the following Riccati equation

\begin{equation}
\label{R2}
W_2^2(\theta)- W_2^{\prime}(\Theta)=\frac{1}{4}
\left\{\frac{k(k+1)}{sin^{2}(\Theta/2)} +
\frac{\lambda(\lambda+1)}{cos^{2}(\Theta/2)}\right\}-E^{(0)}_{2},
\end{equation}
which provides a new superpotential and the ground state energy of $H_2$

\begin{equation}
\label{W2}
W_2(\Theta)=
-\frac{(k+2)}{2} cot(\Theta/2) +\frac{(\lambda+2)}{2}tan(\Theta/2), \quad
E^{(0)}_{2} = \frac{1}{4} (k+\lambda+2)^{2}.
\end{equation}
Thus the eigenfunction associated to the ground state of $H_2$
is given by

\begin{equation}
\label{GS2} \psi^{(0)}_{2}=exp\left\{-\int
W_2(\Theta)d\Theta\right\}\propto
sin^{k+1}(\Theta/2)cos^{\lambda+1}(\Theta/2).
\end{equation}
Hence in analogy with (\ref{A1-}) and (\ref{A1+})
the new intertwining operators are given by

\begin{eqnarray}
\label{A2NN}
A^{\pm}_{2} &&= \pm\frac{d}{d\Theta} -W_2(\Theta)\nonumber\\
 {}&&=\pm\frac{d}{d\Theta}+\frac{(k+1)}{2}cot(\Theta/2) -
\frac{(\lambda+1)}{2}tan(\Theta/2)\nonumber\\
A^{-}_{2}&&= \psi^{(0)}_{2}\left(-\frac{d}{d\Theta}\right)
\frac{1}{\psi^{(0)}_2}, \quad A^{-}_{2}\psi^{(0)}_2= 0.
\end{eqnarray}

Note that the $V_2(\Theta)$ partner potential has a symmetry, viz.,

\begin{equation}
\label{V2}
V_2(\Theta)= \frac{1}{4} \left(\frac{k(k+1)}{sin^{2}(\Theta/2)} +
\frac{\lambda(\lambda+1)}{cos^{2}(\Theta/2)}\right)=
V_1(k\rightarrow k+1,\lambda \rightarrow \lambda+1)
\end{equation}
which is leads to the shape-invariance property (outlined in subsection
III.2) for the first unbroken SUSY potential pair

\begin{eqnarray}
\label{H2}
V_{1-}&&=\frac{1}{4} \left(\frac{k(k-1)}{sin^{2}(\Theta/2)} +
\frac{\lambda(\lambda-1)}{cos^{2}(\Theta/2)}\right)-
\frac{1}{4}(k+\lambda)^2 \nonumber\\
V_{1+}&&= \frac{1}{4} \left(\frac{k(k+1)}{sin^{2}(\Theta/2)} +
\frac{\lambda(\lambda+1)}{cos^{2}(\Theta/2)}\right)-
\frac{1}{4}(k+\lambda)^2 \nonumber\\
{}&&=V_{1-}(k\rightarrow k+1,\lambda \rightarrow \lambda+1)
+(\lambda +k+1).
\end{eqnarray}
 In this case, one can obtain energy eigenvalues and eigenfunctions
   by means of the
  shape-invariance condition. However, we have derived the excited
  state algebraically, by exploiting the Sukumar's method for the
  construction of SUSY hierarchy \cite{Suku85}. Furthermore, note that
  $\psi^{(0)}_{1-}=\psi^{(0)}_{1}$ is normalizable with zero energy for the
  ground state of bosonic sector Hamiltonian $H_{1-}=H_1-E^{(0)}_{1}$ and
  the
  energy eigenvalue for the ground state of fermionic sector Hamiltonian
  $H_{1+}=H_2-E^{(0)}_{1}$ is exactly the first excited state of $H_{1-},$ but
  the eigenfunction $\frac{1}{\psi^{(0)}_{1}}$ is not the ground state of
  $H_{1+},$ for $k>0$ and $\lambda>0.$

Let us again consider the Sukumar's method in order
to find the partner potential of $V_2(\Theta)$ is

\begin{eqnarray}
\label{V3} V_3(\Theta)&&= V_2(\Theta)-2\frac{d^2}{d\Theta^2}\ell
n\psi^{(0)}_2= V_2(\Theta)+ \frac{1}{2}
\left(\frac{(2k+1)}{sin^{2}(\Theta/2)} +
\frac{(2\lambda+1)}{cos^{2}(\Theta/2)}\right)\nonumber\\ {}&&= \frac{1}{4}
\left(\frac{(k+2)(k+1)}{sin^{2}(\Theta/2)} +
\frac{(\lambda+2)(\lambda+1)}{cos^{2}(\Theta/2)}\right)= V_1(k\rightarrow
k+2,\lambda \rightarrow \lambda+2).
\end{eqnarray}
Now one is able to implement the generalization for
nth-member of the hierarchy,  i.e.
the general potential may be written for all integer values of $n,$
viz.,

\begin{eqnarray}
\label{Vn} V_n(\Theta)&&= V_1(\Theta)+\frac{1}{4}
(n-1)\left\{\frac{2k+n-2}{sin^{2}(\Theta/2)} +
\frac{2\lambda+n-2}{cos^{2}(\Theta/2)}\right\}\nonumber\\
{}&&=\frac{k^2-k+k(n-1)+(n-1)(n-2)}{4sin^{2}(\Theta/2)} \nonumber\\
{}&&+\frac{\lambda^2-\lambda+\lambda(n-1)+(n-1)(n-2)}{4cos^{2}(\Theta/2)}
\nonumber\\ {}&&=
\frac{1}{4}\left\{\frac{(k+n-1)(k+n-2)}{sin^{2}(\Theta/2)} +
\frac{(\lambda+n-1)(\lambda+n-2)}{cos^{2}(\Theta/2)}\right\}.
\end{eqnarray}
Note that $V_n(\Theta)=V_1(\Theta; k\rightarrow k+n-1,
 \lambda\rightarrow \lambda+n-1)$ so that the
(n+1)th-member of the hierarchy is given by

\begin{equation}
\label{Hn+}
H_{n+1} = A^{+}_{n+1} A^{-}_{n+1} + E^{(0)}_{n+1},
\quad E^{(0)}_{n+1} = \frac{1}{4}(k+\lambda+2n)^2
\end{equation}
where

\begin{eqnarray}
\label{An+} A^{-}_{n+1} &&=  \psi^{(0)}_{n+1}
\left(-\frac{d}{d\Theta}\right) \frac{1}{\psi^{(0)}_{n+1}} =
\left(A^+_{n+1}\right)^{\dagger}\nonumber\\
\psi^{(0)}_{n+1} &&\propto
sin^{k+n}(\Theta/2)cos^{\lambda+n}(\Theta/2).
\end{eqnarray}
Applying the SUSY hierarchy method (\ref{E60a}), one gets the
nth-excited state of $H_1$ from the ground state of $H_{n+1}$,
as given by

\begin{eqnarray}
\label{pn}
\psi^{(n)}_{1}&&\propto A^{+}_{1} A^{+}_{2}\ldots A^{+}_{n}
sin^{k+n} (\Theta/2) cos^{\lambda+n} (\Theta/2)\nonumber\\
{}&&=\prod_{s=0}^{n-1}
\left[\frac{1}{sin^{k+s}(\Theta/2)cos^{\lambda+s}(\Theta/2)}
\left(\frac{d}{d\Theta}\right)\right.\nonumber\\
{}&&\left. sin^{k+s}(\Theta/2)
cos^{\lambda+s}(\Theta/2)\right]sin^{k+n} (\Theta/2)
cos^{\lambda+n}(\Theta/2)\nonumber\\
{}&&\propto\frac{1}{sen^{k-1}(\Theta/2)cos^{\lambda-1}(\Theta/2)}
\left(\frac{1}{sin(\Theta/2)} \frac{d}{d\Theta}\right)^n
sin^{2(k+n)-1}(\Theta/2) cos^{2(\lambda+n)-1}(\Theta/2)\nonumber\\
&&\propto(1-u)^{\frac{k}{2}} (1+u)^{\frac{\lambda}{2}} \left[(1-u)^{-k
+\frac 12} (1+u)^{-\lambda+\frac 12}
\left(\frac{d^n}{du^n}\right)(1-u)^{k-\frac 12+n} (1+u)^{\lambda-\frac
12+n}\right],
\end{eqnarray}
where $(u= cos \Theta),$ so that the ground state of nth-member of
hierarchy is given by

\begin{equation}
\psi^{(0)}_n \propto sin^{k+n-1} (\Theta/2) cos^{\lambda+n-1} (\Theta/2)
\end{equation}
and the nth-excited state of PTI potential become
\begin{equation}
\label{pn1}
\psi_1^{(n)}(\Theta)\propto sin^{k}(\Theta/2) cos^{\lambda}(\Theta/2)
F\left(-n, n+k+\lambda; k+\frac{1}{2};sin^{2}(\Theta/2)\right),
\end{equation}
which follows on
identification of the square bracketed quantity in (\ref{pn}) with the Jacobi
polynomials (Gradshteyn and Ryzhik \cite{Grad})

$$ J_n^{\left(k-\frac{1}{2},\lambda-\frac{1}{2}\right)}\propto
F\left(-n, n+k+\lambda;k+\frac{1}{2};\frac{1-u}{2}\right).
$$
Here $F$ are known as the confluent hypergeometric functions
which clearly they are in the region of convergency and defined
by \cite{GR}

$$
F(a,b;c;x)=1+\frac{ab}{c}x+\frac{a(a+1)(b(b+1)}{1.2.c(c+1)}x^2+
\frac{a(a+1)(a+2)b(b+1)(b+2)}{1.2.c(c+1)(c+2)}x^3+\cdots
$$
and its derivative with respect to $x$ becomes

$$
\frac{d}{dx}F(a,b;c;x)=\frac{ab}{c}F(a+1,b+1;c+1;x).
$$

The excited state eigenfunctions (\ref{pn1}) here obtained  by the SUSY
algebraic method agree with those given in Fl\"ugge \cite{Flu} using
non-algebraic method. Note that the coupling constants $k$ and
$\lambda$ in above analysis are unquantized.
Besides from Sec. III, Eq. (\ref{Hn+}) and Eq. (\ref{NPT})
we readily find the following energy eigenvalues for the PTI potential

\begin{eqnarray}
\label{EPT} E^{(n)}_{1} &&=E^{(n-1)}_{2}=\cdots=2\alpha^2E^{(0)}_{n+1}
=\frac{\alpha^2}{4}(k+\lambda+2n)^2, \nonumber\\ E^{(n)}_{PT}
&&=2\alpha^2E^{(n)}_1=\frac{\alpha^2}{4}(k+\lambda+2n)^2, \quad
n=0,1,2,\cdots_.
\end{eqnarray}

Let us now point out the existence of various possibilities to
  supersymmetrize the PTI Hamiltonian with broken SUSY, for a finite interval
  $[0, \pi].$ with
  unbroken and broken SUSY. Indeed both ground states do not have zero
  energy, so that when  $k$ and $\lambda$ are in a particular interval
  one can have a broken SUSY, because there such eigenstates are
  also not normalizable.
  In these cases, the  shape invariant procedure is
  not valid but the Sukumar's SUSY hierarchy procedure \cite{Suku85}
  can be applied.
Therefore, we see that other combinations of the $k$ and
  $\lambda$ parameters are also possible to provide distinct
  superpotentials.

\section{NEW SCENARIO OF SUSY QM}

In this Section we apply the SUSY QM for a neutron in interaction with a
static magnetic field of a straight current carrying wire. This system is
described by two-component wave functions, so that the development
considered so far for SUSY QM must be adapted.

The essential reason for the
necessity of modification is due to the Riccati equation may be reduced to
a set of first-order coupled differential equations. In this case the
superpotential is not defined as $W(x)=-\frac{d}{dx}\ell n
\left(\psi_0(x)\right),$ where $\psi_0(x)$ is the two-component
eigenfunction of the ground state. Only in the case of 1-component wave
functions one may write the superpotential in this form. Recently two
superpotentials, energy eigenvalue and the two-component eigenfunction of
the ground state have been found \cite{VGM,RVV01}.

In this Subection we investigate a symmetry between the supersymmetric
Hamiltonian pair ${\bf H}_{\pm}$ for a neutron in an external magnetic
field.
After some transformations on the
original problem
which corresponds to a one-dimensional Schr\"odinger-like equation
associated with the two-component wavefunctions
in cylindrical coordinates, satisfying the following eigenvalue
equation

\begin{equation}
{\bf H}_{\pm}{\bf\Phi}^{(n_\rho,m)}_{\pm}=E^{(n_\rho,m)}_{\pm}
{\bf\Phi}^{(n_\rho,m)}_{\pm}, \quad n_\rho=0,1,2, \cdots,
\end{equation}
where $n_\rho$ is the radial quantum number and $m$ is the orbital
angular momentum eigenvalue
in the z-direction. The two-component energy eigenfunctions are given by

\begin{equation}
\label{wfn}
{\bf\Phi}^{(n_\rho,m)}_{\pm}={\bf\Phi}^{(n_\rho,m)}_{\pm}(\rho,k)=
\left(
\begin{array}{c}
\phi_{1\pm}^{(n_\rho,m)}(\rho,k) \\
\phi_{2\pm}^{(n_\rho,m)}(\rho,k)
\end{array}
\right)
\end{equation}
and the supersymmetric Hamiltonian pair

\begin{eqnarray}
\label{Hpm}
{\bf H}_-&&\equiv {\bf A}^{+}{\bf A^{-}}
=-\mbox{\bf I}\frac{d^2}{d\rho^2}+\left(
\begin{array}{llll}
\frac{m^2-\frac{1}{4}}{\rho^2}+\frac{1}{8(m+1)^2} & \;\;\;\frac{1}{\rho}\\
\frac{1}{\rho} &  \frac{(m+1)^2-\frac{1}{4}}{\rho^2}+\frac{1}{8(m+1)^2}
\end{array}
\right)\nonumber\\
{\bf H}_+&&\equiv {\bf A}^{-}{\bf A^{+}}=
{\bf H}_--[ {\bf A}^{+}, {\bf A^{-}}]_-\nonumber\\
{}&&=-\mbox{\bf I}\frac{d^2}{d\rho^2}+
\hbox{\bf V}_{+},
\end{eqnarray}
where {\bf I} is the 2x2 unit matrix and

\begin{equation}
{\bf A}^{\pm}=\pm\frac{d}{d\rho}+{\bf W}(\rho).
\end{equation}
In this Section we are using the notation of Ref. \cite{RVV01}.
Thus in this
case the Riccati equation in matrix form, becomes

\begin{equation}
\mbox{\bf W}^{\prime}(\rho)+\mbox{\bf W}^2(\rho)= \frac{(m+\frac
12)(m+\frac 12-\sigma_3)}{\rho^2}+ \frac{\sigma_1}{\rho}+\frac{{\bf I}
}{8(m+1)^2},
\end{equation}
where the  two-by-two hermitian superpotential matrix
recently calculated in \cite{RVV01}, which is given by

\begin{equation}
\label{SPW2} {\bf W}(\rho; m)={\bf W}^{\dagger}= (m+\frac 12)({\bf
I}+\sigma_3)\frac{1}{2\rho}+ (m+\frac 32)({\bf
I}-\sigma_3)\frac{1}{2\rho}+\frac{\sigma_1}{2m+2},
\end{equation}
 where $\sigma_1, \sigma_3$ are the well known Pauli matrices.

The hermiticity condition on the superpotential matrix allows us to
construct the following  supersymmetric potential partner

\begin{eqnarray}
\label{Vpm}
{\bf V}_+(\rho;m )&&={\bf V}_--2\mbox{\bf W}^{\prime}(\rho)\nonumber\\
{}&&=\mbox{\bf W}^2(\rho)-\mbox{\bf W}^{\prime}(\rho)\nonumber\\
{}&&=\left(\begin{array}{llll}
\frac{(m+\frac 12)(m+\frac{3}{2})}{\rho^2} & \;\;\;\frac{1}{\rho}\\
\frac{1}{\rho} &  \frac{(m+\frac 12)(m+\frac{7}{2})+2}{\rho^2}
\end{array}
\right)+\frac{{\bf I}}{8(m+1)^2}.
\end{eqnarray}

Note that in this case we have unbroken SUSY because the ground state has
zero energy, viz., $E_-{(0)}=0,$
with the annihilation conditions

\begin{equation}
\label{AC1}
A^- \Phi^{(0)}_- = 0
\end{equation}
and

\begin{equation}
\label{AC2}
A^+ \Phi^{(0)}_+ = 0.
\end{equation}
Due to the fact these eigenfunctions to be of two components
one is not able  to write the superpotential in terms of
them in a similar way of that one-component eigenfunction belonging
to the respective ground state.

Furthermore, we have a symmetry between ${\bf V}_{\pm}(\rho;m).$
Indeed, it is easy to see that

\begin{eqnarray}
{\bf V}_+(\rho; m)&&=\frac{(m+1)^2-\frac 14}{\rho^2}{\bf I}+
(2m+3)\frac{({\bf I}-\sigma_3)}{2\rho^2}+
\frac{\sigma_1}{\rho}+\frac{{\bf I}}{8(m+1)^2}\nonumber\\
{}&&={\bf V}_-(\rho;m\rightarrow m+1)+{\bf R}_m,
\end{eqnarray}
where ${\bf R}_m=-\frac{{\bf I}}{8}(2m+3)(m+1)^{-2}(m+2)^{-2}.$
Therefore, we can find the energy eigenvalue and eigenfunction of the
ground state of ${\bf H}_+$ from those of ${\bf H}_-$
and the resolution spectral of the hierarchy of matrix Hamiltonians
can be achieved in an elegant way via the shape invariance procedure.

\newpage
\section{Conclusions}

We start  considering the Lagrangian formalism for the
construction of one dimension  supersymmetric quantum mechanics
with N=2 SUSY in a non-relativistic context, viz., two Grassmann
variables in classical mechanics and the Dirac canonical
quantization method was considered.

This paper also relies on known connections between the theory of Darboux
operators \cite{darboux} in factorizable essentially isospectral partner
Hamiltonians (often called as supersymmetry in quantum mechanics "SUSY
QM"). The structure of the Lie superalgebra, that incorporates commutation
and anticommutation relations in fact characterizes a new type of a
dynamical symmetry which is SUSY, i.e., a symmetry that converts bosonic
state into fermionic state and vice-versa with the Hamiltonian, one of the
generators of this superalgebra, remaining invariant under such transformations
\cite{textqm,W}. This aspect of it as well reflects in its tremendous
physical content in Quantum Mechanics as it connects different quantum
systems which are otherwise seemingly unrelated.

A general review on the SUSY algebra in quantum mechanics and the procedure
on like to build a SUSY Hamiltonian hierarchy in order of a complete
spectral resolution it was explicitly applied for the  P\"oschl-Teller
potential I. We will now do a more detail discussion for the case of this
problem presents broken SUSY.

It is well known that usual shape invariant procedure \cite{Gend,inv01} is
not applicable for computation energy spectrum of a potential without zero
energy eigenvalue. Recently, the approach implemented with a two-step shape
invariance in order to connect broken and unbroken \cite{Sukhat93} is
considered in connections \cite{Sukhat01}. In these references it is
considered the P\"oschl-Teller potential I (PTPI), showing the types of
shape invariance it possesses. In this work we consider  superpotential
continuous and differentiable that provided us the PTPI SUSY partner with
the nonzero energy eigenvalue for the ground state, a broken SUSY system,
or containing a zero energy for the ground state with unbroken SUSY. We
have presented our own application of the SUSY hierarchy method, which can
also be applied for broken SUSY \cite{Suku85}.
The potential algebras for shape invariance potentials
have been considered in the references \cite{Sukhat01,balan98}.

We have also applied the SUSY QM formalism  for a neutron in
interaction with a static magnetic field of a straight
current carrying wire, which is described by two-component wave
functions, and presented a new scenario in the coordinate representation.
Parts of such an application have also been considered in
\cite{VGM,RVV01}.

Furthermore, we stress that defining $k=-2(m+\frac 12)$ and
$\lambda=2(m+\frac 12),$
where $m$ is angular moment along $z$ axis,
in the PTPI
it is possible to obtain the energy eigenvalue and eigenfunctions
for such a planar physical system
as an example of the 2-dimensional supersymmetic problem in
the momentum representation \cite{Voro}.
We see from distinct superpotentials may be considered distinct
supersymmetrizations of the PTPI potential \cite{pc01r}.

In this article some applications of SUSY QM were not commented. As
examples, the connection between SUSY and the variational and the WKB
methods. In \cite{Lahi,Fred} the reader can find various references about
useful SUSY QM and in improving the old WKB and variational method.
However, the WKB approximations provide us good results for higher states
than for lower ones. Hence if we apply the WKB method in order to calculate
the ground state one obtain a very poor approximation. The $N=2$ SUSY
algebra  and many applications including its connection with the
variational method and supersymmetric WKB have been recently studied in the
literature \cite{Elso94,Dutt95}. Indeed, have been suggested that
supersymmetric WKB method may be useful in studying the deviation of the
energy levels of a quantum system due to the presence of spherically
confining boundary \cite{Dutt95}. There they have observed that the
confining geometry removes the angular-momentum degeneracy of the
electronic energy levels of a free atom. Khare has investigated the
supersymmetric WKB quantization approximation \cite{khare85}, and
Khare-Yarshi have studied the bound state spectrum of two classes of
exactly solvable non-shape-invariant potentials in the SWKB approximation
and shown that it is not exact \cite{Khare89}. A method to obtain wave
function in a uniform semiclassical approximation to SUSY QM has been
applied for the Morse, Rosen-Morse, and anharmonic oscillator potentials
\cite{Hatchell88}. Inomata {\it et al.} have applied the WKB quantization
rule for the isotropic harmonic oscillator in three dimensions, quadratic
potential and the PTPI \cite{inomata93}.

In literature, the SUSY algebra has also been applied to invetigate a
variety of one-parameter families of isospectral SUSY partner potentials
\cite{Nieto84,suku85inv,sukhat89} in non-relativistic quantum mechanics
which are phase-equivalent \cite{Talu}. By phase-equivalent potentials it
is understood that the potentials relate all Hamiltonians which have the
same phase shifts and essentially the same bound-state spectrum.
L\'evai-Baye-Sparenberg have obtained potentials which are phase-equivalent
with the generalised Ginochio potential \cite{levai97}. Nag-Roychoudhury
show that the repeated application of  Darboux's theorem \cite{darboux} for
an isospectral Hamiltonian provides a new potential which can be phase
equivalent. However, such a similar procedure is inequivalent to the usual
approach on Darboux' theorem \cite{roychou95}.

Another important approach is the connection between SUSY QM and the Dirac
equation, so that many authors have considered in their works. For example,
Ui \cite{Ui84} has shown that a Dirac particle coupled to a Gauge field in
three spacetime dimensions possesses a SUSY analogous to Witten's model
\cite{W} and Gamboa-Zanelli \cite{Gamboa88} have discussed the extension to
include non-Abelian Gauge fields, based on the ground-state wavefunction
representation for SUSY QM. The SUSY QM has also been applied for the Dirac
equation of the electron in an attractive central Coulomb field by Sukumar
\cite{Sukud85}, to a massless Dirac particle in a magnetic field by
Huges-Kostelecty-Nieto \cite{Richard86}, and to second-order relativistic
equations, based on the algebra of SUSY by Jarvis-Stedman \cite{Jarvis86}
and for a neutral particle with an anomalous magnetic moment in a central
electrostatic field by Fred {\it et al.} \cite{Fred88} and Semenov
\cite{Seme90}. Beckers {\it et al.} have shown that $2n$ fermionic
variables of the spin-orbit coupling procedure may generate a grading
leading to a unitary Lie superalgebra \cite{Beckers90}. Using the
intertwining of exactly solvable Dirac equations with one-dimensional
potentials, Anderson has shown that a class of exactly solvable potentials
corresponds to solitons of the modified Korteweg de Vries equation
\cite{Ander91}. Njock {\it et al.} \cite{Njock94} have investigated the
Dirac equation in the central approximation with the Coulomb potential, so
that they have derived the SUSY-based quantum defect wave functions from an
effective Dirac equation for a valence that is solvable in the limit of
exact quantum-defect theory. Dahl-Jorgensen have investigated the
relativistic Kepler prblem with emphasis on SUSY QM via Jonson-Lippman
operator \cite{dahl94}. The energy eigenvalues of a Dirac electron in a
uniform magnetic field has been analyzed via SUSY QM by Lee \cite{Lee94}.
The relation between superconductivity and Dirac SUSY has been generalized
to a multicomponent fermionic system by Moreno {\it et al.}
\cite{Moreno95}.

 An interesting quantum system is the so-called Dirac oscillator,
first introduced by Moshinsky-Szczepaniak \cite{Mos89}; its spectral
resolution has been investigated with the help of techniques of  SUSY QM
\cite{Beni90}. The Dirac oscillator with a generalized interaction has been
treated by Casta\~nos {\it et al.} \cite{Castanos91}. In another work,
Dixit {\it et al.} \cite{Dixit92} have considered  the Dirac oscillator
with a scalar coupling whose non-relativistic limit leads to a harmonic
oscillator Hamiltonian plus a $\vec S\cdot \vec{\hat r}$ coupling term. The
wave equation is not invariant under parity. They have worked out a
parity-invariant Dirac oscillator with scalar coupling by doubling the
number of components of the wave function and  using the Clifford algebra
$C\ell_7.$ These works motivate the construction of a new linear
Hamiltonian in terms of the  momentum, position and mass coordinates,
through a set of seven  mutually anticommuting 8x8-matrices yielding a
representation of the Clifford algebra $C\ell_7$. The seven elements of the
Clifford algebra $C\ell_7$ generate the three linear momentum components,
the three position coordinates components and the mass, and their squares
are the 8x8-identity matrix {\bf I}$_{8\hbox{x}8}.$

Recently, the Dirac
oscillator have been approached in terms of a system of two particles
\cite{Mos96} and to Dirac-M\"orsen problem \cite{Alhai} and relativistic
extensions of shape invariant potential classes \cite{inv01}.

Results of
our analysis on the SUSY QM and Dirac equation for a linear potential
\cite{Alhai,Class84} and for the Dirac oscillator via R-deformed
Heisenberg algebra \cite{Jaya,Mik00}, and
the new Dirac oscillator via Clifford algebra $C\ell_7$  are in
preparation.

Crater and Alstine have applied the constraint formalism for
two-body Dirac equations
in the case of general covariant interactions \cite{crater82,crater87},
which has its origin in the work of Galv\~ao-Teitelboim \cite{Galv80}.
This issue and the discussion on the role SUSY
plays to justify the origin of the constraint
have recently been reviewed by Crater and Alstine
\cite{crater00}.

In \cite{Robi97}, Robnit purposes to  generate the superpotential in terms
of arbitrary higher excited
eigenstates, but there the whole formalism of the 1-component SUSY
QM is needed of an accurate analysis due to the nodes from some
excited eigenstates. Results of such investigations  will be
reported separately.

Now let us point out various other interesting applications of the
superymmetric quantum mechanics, for example, the extension dynamical
algebra of the n-dimensional harmonic oscillator with one second-order
parafermionic degree of freedom by Durand-Vinet \cite{Durand90}. Indeed,
these authors have shown that the parasupersymmetry in non-relativistic
quantum mechanics generalize the standard SUSY transformations
\cite{Durand89}. The parasupersymmetry has also been analyzed in the
following references
\cite{parasu91,parasu1,parasu2,parasu3,parasu94,parasu4}.

Other applications of SUSY QM may be found in
\cite{sourlas79,nicolai80,Gen85,Swamy85,sourlas85,graham85,Roy86,Raghu86,jafee87,jafee88,VWW90,cerve91,abe91,Fry93,truax93,Ziem94,junker94,rosu95,NR95,Samso95,Lee95,Suku95,Rosu96,gates96,FF97,Chaturvet98,Plyus98,Roy98,brink98,AIN99,WS99,Hull99,Milos00,ER01,rosu01,Shif00Ritz,Hidea01,Masa01,Tanaka01,Mota01,Seok01,ghosh01}.
All realizations of SUSY QM in these works is based in the Witten's model
\cite{W}. However, another approach on the SUSY has been implemented in
classical and quantum mechanics. Indeed, a $N$=4 SUSY representation in
terms of three bosonic and four fermionic variables transforming as a
vector and complex spinor of rotation group O(3) has been proposed, based
on the supercoordinate construction of the action \cite{pash91}. However,
the superfield SUSY QM with 3 bosonic and 4 fermionic fields was first
described by Ivanov-Smilga \cite{ivanov91}.

It is well known that $N$=4
SUSY is the largest number of extended SUSY for which a superfield
(supercoordinate) formalism is known. However, using components fields and
computations the $N>4$ classification of N-extended SUSY QM models have
been implemented via irreducible multiplets of their representation by
Gates {\it et al.} \cite{gates95}. Recently, Pashnev-Toppan have also shown
that all irreducible multiplets of representation of $N$ extended SUSY are
associated to fundamental short multiplets in which all bosons and all
fermions are accommodated into just two spin states
\cite{Toppan01}.

$N=4$ supersymmetric quantum mechanics many-body systems
in terms of Calogero models and $N=4$
superfield formulations have been investigated by Wyllard \cite{wyllard00}.
This Ref. and the N=4 superfield formalism
used there are actually based on the paper \cite{ivanov89}.

SUSY $N$=4 in terms of the dynamics of a spinning particle in a curved
background has also been described using the superfield
formalism \cite{rive00}.
There are a few more of works where SUSY QM in higher dimensions is
investigated \cite{pashnev00,beluc01,nerse01}.
However, the Ref. \cite{pashnev00} is a further extension of the
results obtained  earlier in the basic paper \cite{ivanov91b}.

The paper of Claudson-Halpern, \cite{halpern84}, was the first to give the
$N=4$ and $N=16$ SUSY gauge quantum mechanics, the latter now called
M(atrix) theory \cite{halpern98}.

\vspace{1cm}

\centerline {\bf {ACKNOWLEDGMENTS}}

\vspace{1.0cm}

The author is grateful to J. Jayaraman, whose advises and encouragement
were useful and also for having made the first fruitful discussions on SUSY
QM. Thanks are also due to J. A. Helayel Neto for teaching me the
foundations of SUSY QM and for the hospitality at CBPF-MCT. This research
was supported in part by CNPq (Brazilian Research Agency). We wish to thank
the staff of the CBPF and DCEN-CFP-UFCG for the facilities. The author
would also like to thank M. Plyushchay, Erik D'Hoker, H. Crater,
B. Bagchi, S. Gates, A. Nersessian, J. W. van Holten, C. Zachos, Halpern,
Gangopadhyaya, Fernandez,  Azc\'arraga, Binosi,
Wimmer, Rabhan, Ivanov and Zhang for the
kind interest in pointing out relevant references on the subject of this
paper. The author is also grateful to A. N. Vaidya, E. Drigo Filho, R. M.
Ricotta, Nathan Berkovits, A. F. de Lima, M. Teresa C. dos Santos Thomas,
R. L. P. Gurgel do Amaral, M. M. de Souza, V. B. Bezerra, P. B. da Silva
Filho, J. B. da Fonseca Neto, F. Toppan, A. Das
and Barcelos for the encouragement
and interesting discussions.

\end{document}